\documentclass[11pt]{article}

\usepackage{authblk}
\usepackage{booktabs}
\usepackage{color}
\usepackage{ulem}
\usepackage{siunitx}
\usepackage{natbib}
\usepackage{enumitem}
\usepackage{amsmath,amssymb,amsfonts,amsthm}
\usepackage{mathtools}
\usepackage{mathabx}
\usepackage{algorithm}
\usepackage{algpseudocode}
\usepackage{setspace}
\usepackage{url}
\usepackage[a4paper,margin=20mm]{geometry}
\usepackage{caption}

\DeclareMathSymbol{\leq}{\mathrel}{symbols}{"14}
\DeclareMathSymbol{\geq}{\mathrel}{symbols}{"15}

\newcommand{\corraddress}[1]{\def\thecorraddress{#1}}
\newcommand{\corremail}[1]{\def\thecorremail{#1}}

\makeatletter
\newcommand{\printcorrespondence}{%
  \begingroup
  \vspace{-3em}
  \begin{center}\small
    \textbf{Correspondence:} \thecorraddress\\
    \textbf{Email:} \texttt{\thecorremail}
  \end{center}
  \endgroup
}
\g@addto@macro\maketitle{\par\vspace{-0.5em}\printcorrespondence}
\makeatother

\title{\fontsize{20pt}{24pt}\selectfont Preemptive Ensemble Forecast Sensitivity to Observations}
\author[1]{Fumitoshi Kawasaki}
\author[2,3,4]{Shunji Kotsuki}
\affil[1]{Graduate School of Science and Engineering, Chiba University, Chiba, Japan}
\affil[2]{Center for Environmental Remote Sensing, Chiba University, Chiba, Japan}
\affil[3]{Institute for Advanced Academic Research, Chiba University, Chiba, Japan}
\affil[4]{Research Institute of Disaster Medicine, Chiba University, Chiba, Japan}
\corraddress{Fumitoshi Kawasaki, Graduate School of Science and Engineering,\\Chiba University, 1-33, Yayoi-cho, Inage-ku, Chiba-shi, Chiba, 263-8522, Japan.}
\corremail{fkawasaki@chiba-u.jp}
\date{}

\begin{document}
    \maketitle
    \begin{abstract}
        \noindent Estimating the impact of observations newly added to an existing numerical weather prediction (NWP) system requires reintegrating the ensemble forecast from the analysis that assimilates the additional observations, which is laborious and computationally expensive.
We therefore propose preemptive EFSO (PEFSO; preemptive ensemble forecast sensitivity to observations), which estimates observation impact without model reintegration by invoking a stronger tangent-linear approximation than that used in EFSO.
The estimated impact, however, is merely a scalar quantity at a specified verification time, so it is also valuable to obtain the ensemble forecast in which the remaining additional observations are assimilated after denying the detrimental ones.
We refer to this procedure as ADD-SEL, but carrying it out requires reintegration.
We therefore propose two methods for updating the ensemble forecast without reintegration as approximations to ADD-SEL, collectively termed ADD-SEL-PRE: one recomputes the ensemble transform matrix (i.e., the Kalman gain), and the other modifies only the innovations at lower cost.
Experiments with the Lorenz-96 model examine whether both can be approximated without reintegration.
PEFSO estimates observation impact comparable to that obtained by EFSO within the range where the tangent-linear approximation remains valid, even under more practical conditions: an ensemble size of 10, the analysis as the reference state, and only the additional observations available.
When additional observations are assimilated after denying detrimental ones, both ADD-SEL-PRE methods update the ensemble forecast, giving forecast errors comparable to those from ADD-SEL, particularly for lead times up to two days, where the tangent-linear approximation is expected to hold.
    \end{abstract}
    \section{Introduction}\label{sec:introduction}
Data assimilation is a technique for estimating statistically more accurate analyses by combining forecasts with observations, and it plays an essential role in numerical weather prediction (NWP).
With recent technological advances, novel observing methods such as phased array weather radar \citep{zrnic_2007,yoshikawa_2013,ushio_2015}, weather drone \citep{inoue_2022}, saildrone \citep{zhang_2023c}, and Global Sounding Balloon \citep{spisak_2026} have been developed, and the number of observations available for data assimilation in NWP systems has increased dramatically.
However, not all newly available observations are beneficial to NWP systems.
Some of them may even be detrimental for various reasons, such as malfunctioning instruments.
Therefore, it is necessary to assess whether each newly obtained observation is beneficial or not, and if so, how beneficial it is.

To evaluate the observation impact on NWP systems, observing system experiments (OSEs; also referred to as data denial experiments) have traditionally been employed as the basic approach \citep[e.g.,][]{kelly_2007}.
In OSEs, however, assimilation--forecast cycles are required both with and without each observation subset under evaluation; hence such assessments have become increasingly costly, in terms of both human effort and computational resources, for modern, highly sophisticated NWP systems.
As a more computationally efficient approach for estimating observation impact, the forecast sensitivity to observations \citep[FSO;][]{langland_2004} was subsequently proposed.
The formulation of \citet{langland_2004} is applicable to variational data assimilation systems employing an adjoint model, and its effectiveness has been demonstrated at operational NWP centers \citep[e.g.,][]{gelaro_2009a,cardinali_2009}.
Furthermore, the ensemble FSO \citep[EFSO;][]{liu_2008c,li_2010,kalnay_2012}, which is based on the ensemble Kalman filter \citep[EnKF;][]{evensen_1994} and estimates observation impact without an adjoint model, was developed and has since been widely used as a powerful tool for observation impact estimation.
For example, \citet{ota_2013a} implemented EFSO in the global EnKF system of the National Centers for Environmental Prediction (NCEP) and demonstrated that EFSO functions well. 
In addition, \citet{sommer_2014} applied the EFSO to a convective-scale LETKF system and showed that the estimated observation impacts were in reasonable agreement with the results of data denial experiments.
While EFSO can estimate observation impacts at a lower computational cost than OSEs, evaluating the impact of newly available observations requires assimilating those observations and then rerunning the ensemble forecast from the resulting analysis.
This requirement can pose a substantial obstacle to observation impact estimation. 
Consider, for instance, estimating the impact of newly available observations on the Meso-scale Ensemble Prediction System (MEPS) of the Japan Meteorological Agency (JMA). 
Although MEPS forecast and analysis products are available from JMA, an environment for running MEPS itself is generally not provided. 
Even if such an environment were available, considerable effort would be required to understand the details of the model and the data assimilation system and to implement and execute them. 
Such constraints and difficulties are by no means specific to MEPS but are common to many operational NWP systems.

In this study, we therefore propose Preemptive EFSO (PEFSO), a method for estimating observation impact more readily when additional observations become available, without requiring reintegration of the NWP model.
PEFSO approximates EFSO based on the Preemptive Forecast \citep{etherton_2007} and its derivative, Ultra Rapid Data Assimilation \citep[URDA:][]{potthast_2018,kawasaki_2025}. 
In PEFSO, an ensemble transform matrix is computed with the Ensemble Transform Kalman Filter \citep[ETKF:][]{bishop_2001} or the Local Ensemble Transform Kalman Filter \citep[LETKF:][]{hunt_2007}, and the observation impact is estimated approximately by making use of the existing ensemble forecast, thereby avoiding model reintegration.
EFSO, however, yields only a scalar measure of the forecast error reduction at a prescribed verification time.
The actual forecast obtained by assimilating only those observations identified as beneficial therefore provides information that EFSO cannot, including spatial structures and lead-time dependence.
Obtaining such forecasts, however, requires reintegrating the ensemble forecast from the analysis that assimilates the additional observations.
Following the same idea of avoiding model reintegration, we therefore introduce two methods for approximately obtaining the forecast ensemble at each lead time.
The first recomputes the ensemble transform matrix (i.e., the Kalman gain) based on URDA, whereas the second is a low-cost method that modifies only the innovation, following the approach of \citet{ota_2013a,hotta_2017,chen_2019}.
The scientific question of this study is thus whether the observation impact of additional observations can be estimated and the ensemble forecast updated with sufficient accuracy without model reintegration.
It should be noted, however, that such estimation of observation impact and updating of the ensemble forecast without model reintegration are not intended to replace definitive evaluations such as conventional OSEs or EFSO; rather, they are positioned as a preliminary assessment preceding them.
Nevertheless, the ability to estimate the influence of new observations rapidly and at low cost is of considerable practical significance and is expected to contribute to further promoting observational research.

In this study, numerical experiments are conducted using the 40-variable Lorenz 96 model \citep{lorenz_1996,lorenz_1998a} to verify the effectiveness of the proposed method.
First, to isolate the influence of the approximations introduced in PEFSO and to clarify how it varies with ensemble size, reference state, and verification time, EFSO and PEFSO are compared under full observation.
Next, we investigate the behavior of EFSO and PEFSO in a situation where additional observations are available.
Specifically, the observations at grid points 1--32 are regarded as conventional observations, and assimilation--forecast cycles are run in advance to obtain the analyses and forecasts.
Then, assuming that additional observations newly become available at grid points 33--40, EFSO and PEFSO are compared.
Furthermore, when the remaining observations are assimilated after denying the detrimental ones among the additional observations, we compare the ensemble forecast obtained by reintegrating the forecast model with that updated by the approximate methods.

The remainder of this paper is organized as follows. 
Section \ref{sec:methodology} describes the formulations of EFSO and PEFSO. 
Section \ref{sec:experiment} presents the design of a series of experiments using the Lorenz 96 model. 
Section \ref{sec:result} shows the experimental results, and section \ref{sec:discussion} provides a discussion. 
Finally, section \ref{sec:conclusion} presents the conclusions.
    \section{Methodology}\label{sec:methodology}
\subsection{Observation impact estimation for existing observations}
We first describe the formulations of EFSO and PEFSO for the case in which only the existing observations are assimilated. 
In this setting, since no additional observations are assumed, EFSO can be applied directly once the forecast initialized at time $0$ is available; hence the advantage of PEFSO is not exploited, and this is not the setting that PEFSO is originally intended for.
We nevertheless consider this situation in this study, because the difference between EFSO and PEFSO lies solely in the additional tangent linear approximation introduced in PEFSO, which allows a pure comparison between them.
It also serves as preparation for the formulation with additional observations described in the following subsection.

\subsubsection{EFSO for existing observations} \label{sec:method_efso_general}
\begin{figure}[H]
    \centering
    \includegraphics[width=14.5cm]{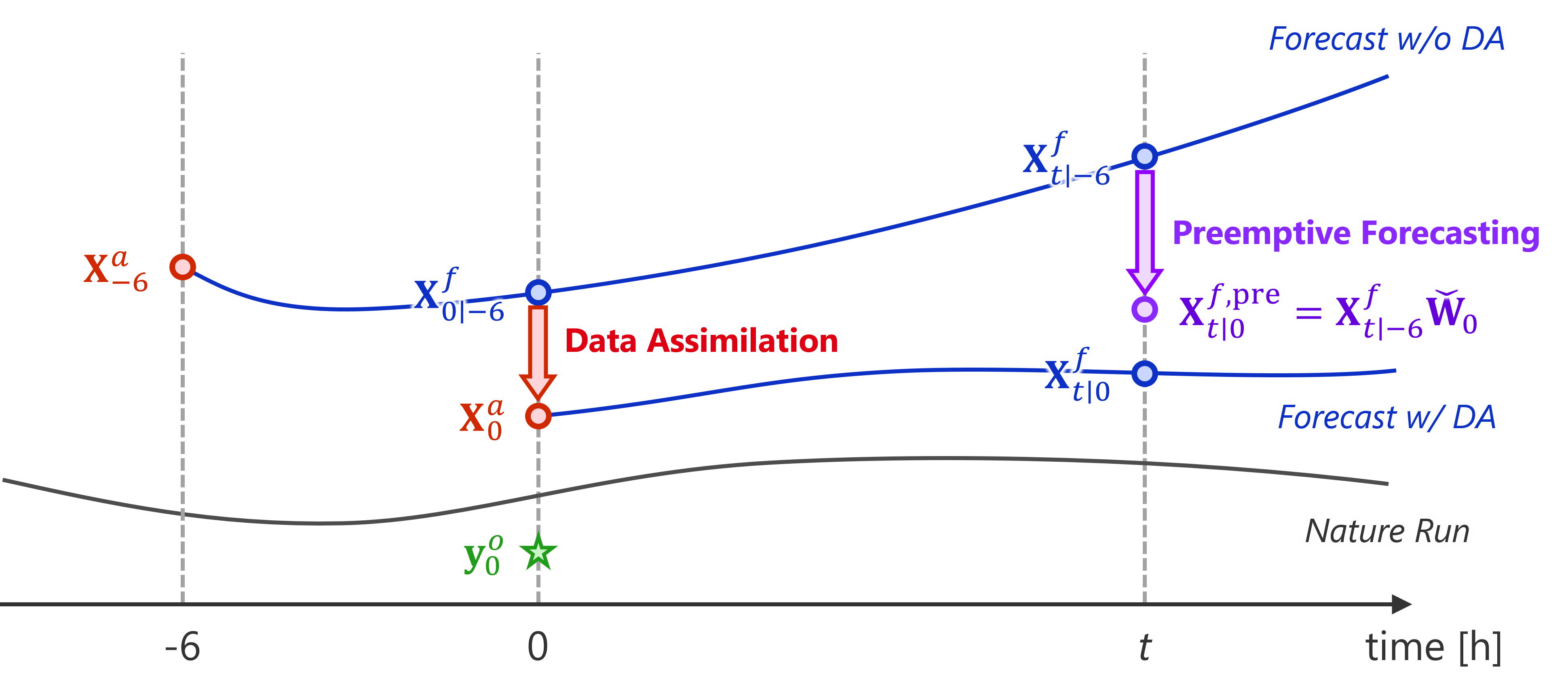}
    \caption{Schematic image of the Ensemble Forecast Sensitivity to Observations (EFSO) and Preemptive EFSO (PEFSO) for existing observations.}
    \label{fig:concept_efso_all}
\end{figure}

EFSO evaluates how much each observation assimilated at the analysis time contributes to reducing the forecast error at a verification time.
A schematic of EFSO is shown in Fig.~\ref{fig:concept_efso_all}.
Hereafter, following common practice, the time of the observations to be diagnosed is denoted by $0$, the preceding analysis time by $-6$, and the verification time of the forecast error by $t$.
As illustrated in Fig.~\ref{fig:concept_efso_all}, the forecast ensembles $\mathbf{X}^{f}_{t\mid-6}$ and $\mathbf{X}^{f}_{t\mid0}$, both valid at time $t$ but initialized at times $-6$ and $0$, differ only in whether the observations $\mathbf{y}^{o}_{0}$ at time $0$ have been assimilated.
The forecast errors of the forecasts initialized at times $-6$ and $0$ and verified at time $t$ are given by
\begin{gather}
    \mathbf{e}_{t\mid-6}^f=\overline{\mathbf{x}}^{f}_{t\mid-6}-\mathbf{x}^{\text{ref}}_{t},\\
    \mathbf{e}_{t\mid0}^f=\overline{\mathbf{x}}^{f}_{t\mid0}-\mathbf{x}^{\text{ref}}_{t},
\end{gather}
where $\overline{\mathbf{x}}^{f}_{t\mid-6}$ and $\overline{\mathbf{x}}^{f}_{t\mid0}$ denote the ensemble means of $\mathbf{X}^{f}_{t\mid-6}$ and $\mathbf{X}^{f}_{t\mid0}$, respectively, and $\mathbf{x}^{\text{ref}}_{t}$ is the reference state at the verification time $t$. 
In this study, we adopt as the reference state $\mathbf{x}^{\text{ref}}_{t}$ either the nature run $\mathbf{x}^{\text{tru}}_{t}$, which is the idealized choice, or the analysis ensemble mean $\overline{\mathbf{x}}^{a}_{t}$, which is the more realistic one.
Following the formulation of \citet{kalnay_2012}, the total observation impact in EFSO is then expressed as
\begin{align}
    \Delta e^{2}_{t}&=\left(\mathbf{e}^{f}_{t\mid0}\right)^{\top}\mathbf{C}\mathbf{e}^{f}_{t\mid0}-\left(\mathbf{e}^{f}_{t\mid-6}\right)^{\top}\mathbf{C}\mathbf{e}^{f}_{t\mid-6} \notag\\
    &=\left(\mathbf{e}^{f}_{t\mid0}-\mathbf{e}^{f}_{t\mid-6}\right)^{\top}\mathbf{C}\left(\mathbf{e}^{f}_{t\mid0}+\mathbf{e}^{f}_{t\mid-6}\right) \notag\\
    &\approx\left(\mathbf{d}^{o-f}_{0}\right)^{\top}\mathbf{R}^{-1}_{0}\mathbf{Y}^{a}_{0}\left(\mathbf{Z}^{f}_{t\mid0}\right)^{\top}\mathbf{C}\left(\mathbf{e}^{f}_{t\mid0}+\mathbf{e}^{f}_{t\mid-6}\right). \label{eq:efso_all}
\end{align}
Here, $\mathbf{C}$ is a positive definite matrix defining the forecast error norm, $\mathbf{R}_{0}$ is the observation error covariance matrix, and $\mathbf{d}^{o-f}_{0}\coloneqq\mathbf{y}^{o}_{0}-H\left(\overline{\mathbf{x}}^{f}_{0\mid-6}\right)$ is the innovation, with $H$ denoting the observation operator. 
Furthermore, letting $\delta\mathbf{X}^{f}_{0\mid-6}$ and $\delta\mathbf{X}^{a}_{0}$ be the forecast and analysis ensemble perturbations, $\mathbf{H}$ the Jacobian of the observation operator, and $m$ the ensemble size, we define $\mathbf{Z}^{f}_{0\mid-6}\coloneqq\delta\mathbf{X}^{f}_{0\mid-6}/\sqrt{m-1}$, $\mathbf{Z}^{a}_{0}\coloneqq\delta\mathbf{X}^{a}_{0}/\sqrt{m-1}$, and $\mathbf{Y}^{a}_{0}\coloneqq\mathbf{H}\mathbf{Z}^{a}_{0}$.
Note that the approximation in Eq.~\eqref{eq:efso_all} arises from the tangent linear approximation around $\overline{\mathbf{x}}^{f}_{t\mid-6}$ and from the ensemble approximation.
Since the sensitivity vector is given by
\begin{equation}
    \frac{\partial\left(\Delta e^{2}_{t}\right)}{\partial\mathbf{d}^{o-f}_{0}}\approx\mathbf{R}^{-1}_{0}\mathbf{Y}^{a}_{0}\left(\mathbf{Z}^{f}_{t\mid0}\right)^{\top}\mathbf{C}\left(\mathbf{e}^{f}_{t\mid0}+\mathbf{e}^{f}_{t\mid-6}\right),
\end{equation}
Eq.~\eqref{eq:efso_all} can be interpreted as the inner product of the innovation $\mathbf{d}^{o-f}_0$ and the sensitivity vector $\partial\left(\Delta e^{2}_{t}\right)/\partial\mathbf{d}^{o-f}_{0}$:
\begin{align}
    \Delta e^{2}_{t}&\approx \left\langle \mathbf{d}^{o-f}_{0},\frac{\partial\left(\Delta e^{2}_{t}\right)}{\partial\mathbf{d}^{o-f}_{0}}\right\rangle \notag\\
    &=\sum^{p}_{i=1}\left[\mathbf{d}^{o-f}_{0}\right]_{i}\left[\frac{\partial\left(\Delta e^{2}_{t}\right)}{\partial\mathbf{d}^{o-f}_{0}}\right]_{i},
\end{align}
where $p$ is the dimension of the observation space.
Consequently, the total observation impact $\Delta e^{2}_{t}$ can be decomposed into the impacts of the individual observations, and the observation impact $\Delta e^{2}_{t, l}$ associated with the $l$-th observation $y^{o}_{0,l}$ is expressed as
\begin{align}
    \Delta e^{2}_{t,l}&\approx\left[\mathbf{d}^{o-f}_{0}\right]_{l}\left[\frac{\partial\left(\Delta e^{2}_{t}\right)}{\partial\mathbf{d}^{o-f}_{0}}\right]_{l} \notag\\
    &=\left[\mathbf{d}^{o-f}_{0}\right]_{l}\left[\mathbf{R}^{-1}_{0}\mathbf{Y}^{a}_{0}\left(\mathbf{Z}^{f}_{t\mid0}\right)^{\top}\mathbf{C}\left(\mathbf{e}^{f}_{t\mid0}+\mathbf{e}^{f}_{t\mid-6}\right)\right]_{l}. \label{eq:efso_individual}
\end{align}
Note that a negative value of the EFSO impact indicates a contribution to the reduction of the forecast error.

It is known that when the ensemble size $m$ is much smaller than the dimension of the model space $n$, localization is required for EFSO as well as for the EnKF \citep{kalnay_2012,ota_2013a}. 
When $\mathbf{C}$ is a diagonal matrix, the observation impact associated with the $l$-th observation on the forecast at the $j$-th grid point is expressed as
\begin{equation}
    \left(\Delta e^{2}_{t,l,j}\right)^{\text{loc}}\approx\left[\mathbf{d}^{o-f}_{0}\right]_{l}\left(L_{l,j}\left[\mathbf{R}^{-1}_{0}\right]_{ll}\left[\mathbf{Y}^{a}_{0}\right]_{l}\left[\mathbf{Z}^{f}_{t\mid0}\right]^{\top}_{j}\left[\mathbf{C}\right]_{jj}\left[\mathbf{e}^{f}_{t\mid0}+\mathbf{e}^{f}_{t\mid-6}\right]_{j}\right), \label{eq:efso_loc}
\end{equation}
where $L_{l,j}$ is the localization function for the $l$-th observation and the $j$-th grid point.
However, since the influence of an observation is advected as the forecast proceeds, localization for EFSO should take this time evolution into account.
\citet{kalnay_2012} therefore proposed a method in which the localization center is shifted according to the forecast lead time up to the verification time.
In this study, we refer to this method as advective localization.

It should be noted that, as pointed out by \citet{hotta_2017}, the observation impact estimated by EFSO should not be interpreted as the direct impact of the observation itself; strictly speaking, it measures the impact of the innovation associated with that observation.
Accordingly, even if the observation impact estimated by EFSO is positive (i.e., detrimental), the observation is not necessarily faulty; deficiencies in the NWP model or in the data assimilation process may instead be involved through the background field.

\subsubsection{PEFSO for existing observations}
As shown in Fig.~\ref{fig:concept_efso_all}, in addition to the ensemble forecast from time $-6$ to time $t$, EFSO requires an ensemble forecast valid at time $t$, obtained by reintegrating the forecast model from the analysis ensemble at time $0$.
In contrast, according to the formulation of URDA \citep{kawasaki_2025}, once the forecast ensemble initialized at time $-6$ and integrated up to time $t$ together with the observations at time $0$ is available, the preemptive forecast $\mathbf{X}^{f,\text{pre}}_{t\mid0}$ (hereafter the superscript ``$\text{pre}$'' denotes preemptive) can be obtained under the tangent linear approximation
\begin{equation}
    \mathbf{X}^{f}_{t\mid0}\approx\mathbf{X}^{f,\text{pre}}_{t\mid0}=\mathbf{X}^{f}_{t\mid-6}\widecheck{\mathbf{W}}_{0}. \label{eq:urda_all}
\end{equation}
Here, $\widecheck{\mathbf{W}}_{0}$ is the ensemble transform matrix, given by
\begin{gather}
    \widecheck{\mathbf{W}}_{0}=\frac{1}{\sqrt{m-1}}\mathbf{w}_{0}\mathbf{1}^{\top}+\mathbf{W}_{0},\\
    \mathbf{w}_{0}=\widetilde{\mathbf{P}}^{a}_{0}\left(\mathbf{Y}^{f}_{0\mid-6}\right)^{\top}\mathbf{R}^{-1}_{0}\mathbf{d}^{o-f}_{0},\\
    \mathbf{W}_{0}=\left(\widetilde{\mathbf{P}}^{a}_{0}\right)^{1/2},
\end{gather}
where $\widetilde{\mathbf{P}}^{a}_{0}\coloneqq\left[\mathbf{I}+\left(\mathbf{Y}^{f}_{0\mid-6}\right)^{\top}\mathbf{R}^{-1}_{0}\mathbf{Y}^{f}_{0\mid-6}\right]^{-1}$ and $\mathbf{Y}^{f}_{0\mid-6}\coloneqq\mathbf{H}\mathbf{Z}^{f}_{0\mid-6}$, $\mathbf{1}$ is the all-ones vector, and $\mathbf{I}$ is the identity matrix.
Furthermore, decomposing Eq.~\eqref{eq:urda_all} into its mean and the perturbations yields
\begin{gather}
    \overline{\mathbf{x}}^{f}_{t\mid0}\approx\overline{\mathbf{x}}^{f,\text{pre}}_{t\mid0}=\overline{\mathbf{x}}^{f}_{t\mid-6}+\mathbf{Z}^{f}_{t\mid-6}\mathbf{w}_{0},\\
    \delta\mathbf{X}^{f}_{t\mid0}\approx\delta\mathbf{X}^{f,\text{pre}}_{t\mid0}=\delta\mathbf{X}^{f}_{t\mid-6}\mathbf{W}_{0}.
\end{gather}
It then follows that $\mathbf{Z}^{f,\text{pre}}_{t\mid0}=\mathbf{Z}^{f}_{t\mid-6}\mathbf{W}_{0}$ and $\mathbf{e}^{f,\text{pre}}_{t\mid0}+\mathbf{e}^{f}_{t\mid-6}=2\mathbf{e}^{f}_{t\mid-6}+\mathbf{Z}^{f}_{t\mid-6}\mathbf{w}_{0}$.
Accordingly, using the preemptive forecast, Eq.~\eqref{eq:efso_all} can be rewritten as
\begin{align}
    \Delta e^{2}_{t}&\approx\left(\mathbf{d}^{o-f}_{0}\right)^{\top}\mathbf{R}^{-1}_{0}\mathbf{Y}^{a}_{0}\left(\mathbf{Z}^{f,\text{pre}}_{t\mid0}\right)^{\top}\mathbf{C}\left(\mathbf{e}^{f,\text{pre}}_{t\mid0}+\mathbf{e}^{f}_{t\mid-6}\right) \notag\\
    &=\left(\mathbf{d}^{o-f}_{0}\right)^{\top}\mathbf{R}^{-1}_{0}\mathbf{Y}^{a}_{0}\left(\mathbf{Z}^{f}_{t\mid-6}\mathbf{W}_{0}\right)^{\top}\mathbf{C}\left(2\mathbf{e}^{f}_{t\mid-6}+\mathbf{Z}^{f}_{t\mid-6}\mathbf{w}_{0}\right) \label{eq:pefso_all}.
\end{align}
Note that Eq.~\eqref{eq:pefso_all} invokes a stronger tangent linear approximation than Eq.~\eqref{eq:efso_all}.

In the same manner, the localized observation impact of the $l$-th observation on the forecast at the $j$-th grid point is formulated as follows
\begin{align}
    \left(\Delta e^{2}_{t,l,j}\right)^{\text{loc}}&\approx\left[\mathbf{d}^{o-f}_{0}\right]_{l}\left(L_{l,j}\left[\mathbf{R}^{-1}_{0}\right]_{ll}\left[\mathbf{Y}^{a}_{0}\right]_{l}\left[\mathbf{Z}^{f,\text{pre}}_{t\mid0}\right]^{\top}_{j}\left[\mathbf{C}\right]_{jj}\left[\mathbf{e}^{f,\text{pre}}_{t\mid0}+\mathbf{e}^{f}_{t\mid-6}\right]_{j}\right) \notag\\
    &=\left[\mathbf{d}^{o-f}_{0}\right]_{l}\left\{ L_{l,j}\left[\mathbf{R}^{-1}_{0}\right]_{ll}\left[\mathbf{Y}^{a}_{0}\right]_{l}\left(\left[\mathbf{Z}^{f}_{t\mid-6}\right]_{j}\mathbf{W}^{\text{loc}}_{0,j}\right)^{\top}\left[\mathbf{C}\right]_{jj}\left(\left[2\mathbf{e}^{f}_{t\mid-6}\right]_{j}+\left[\mathbf{Z}^{f}_{t\mid-6}\right]_{j}\mathbf{w}^{\text{loc}}_{0,j}\right)\right\} \label{eq:pefso_all_loc}
\end{align}
where $\mathbf{w}^{\text{loc}}_{0,j}$ and $\mathbf{W}^{\text{loc}}_{0,j}$ are those localized at the $j$-th grid point. 
That is, it should be noted that localization in PEFSO requires the preemptive forecast to be localized as well.

\subsection{Observation impact estimation for additional observations}
This subsection addresses the situation in which new observations become available for an existing NWP system, that is, the situation for which PEFSO is originally intended. 
Here we consider not only the case in which only the additional observations are available, but also the case in which all observations, both the existing and the additional ones, are available.

\subsubsection{EFSO for additional observations}
\begin{figure}[H]
    \centering
    \includegraphics[width=14.5cm]{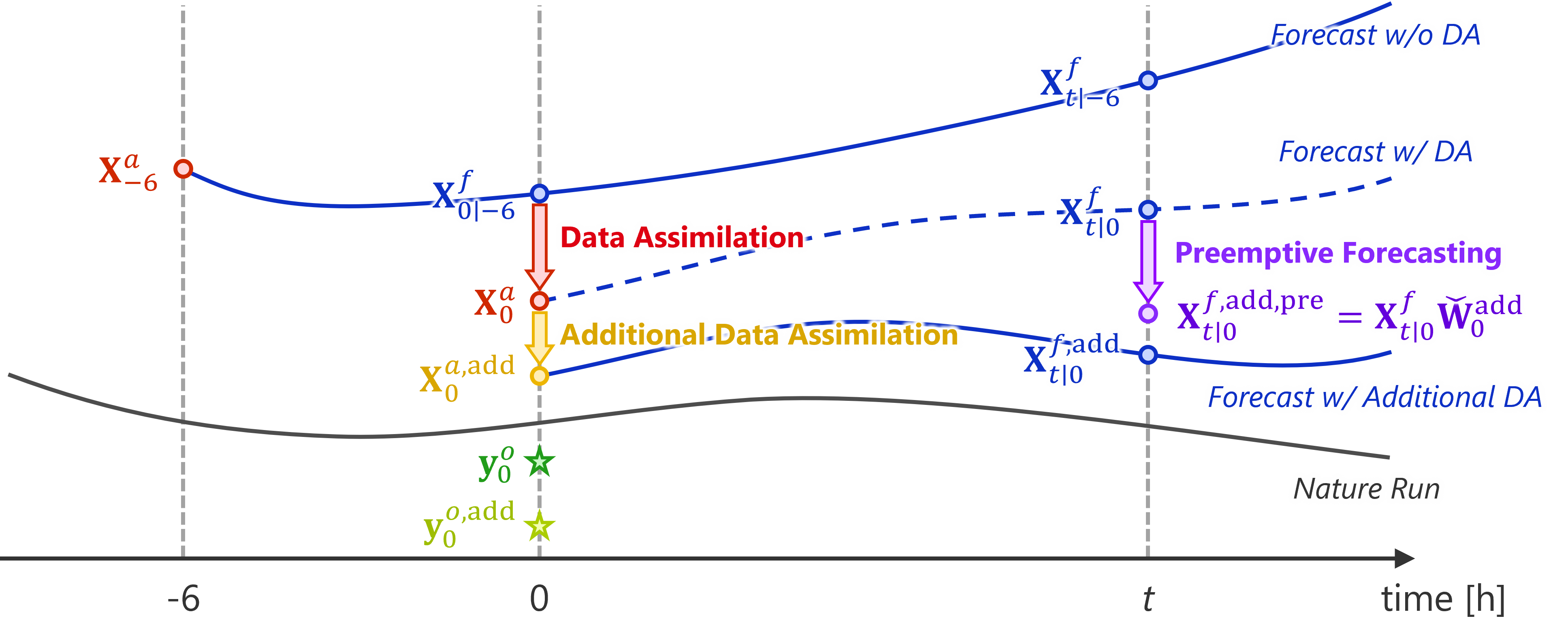}
    \caption{As in Fig.~\ref{fig:concept_efso_all}, but for the case for additional observations.}
    \label{fig:concept_efso_add}
\end{figure}

Figure~\ref{fig:concept_efso_add} shows the situation considered here: in addition to the existing observations $\mathbf{y}^{o}_{0}$ that have been used in the NWP system, new additional observations $\mathbf{y}^{o,\text{add}}_{0}$ become available, and the observation impact is estimated by EFSO.
Here, we assume that the analysis ensembles $\mathbf{X}^{a}_{-6}$ and $\mathbf{X}^{a}_{0}$ at times $-6$ and $0$, together with the forecast ensembles $\mathbf{X}^{f}_{t\mid-6}$ and $\mathbf{X}^{f}_{t\mid0}$, all computed with the existing observations $\mathbf{y}^{o}_{0}$, are available.
This assumption is realistic because such analysis and forecast ensemble products are available in many NWP systems, such as MEPS.
Ideally, the analysis ensemble $\mathbf{X}^{a,\text{add}}_{0}$ should be obtained by assimilating the existing observations $\mathbf{y}^{o}_{0}$ and the additional observations $\mathbf{y}^{o,\text{add}}_{0}$ together.
However, the existing observations $\mathbf{y}^{o}_{0}$ used in the NWP system are not always accessible.
Moreover, reassimilating the existing observations $\mathbf{y}^{o}_{0}$, which have already been assimilated once, entails a wasteful computational cost.
Therefore, in this study, the analysis ensemble $\mathbf{X}^{a,\text{add}}_{0}$ is computed as
\begin{equation}
    \mathbf{X}^{a,\text{add}}_{0}\approx\mathbf{X}^{a}_{0}\widecheck{\mathbf{W}}^{\text{add}}_{0}, \label{eq:analysis_add}
\end{equation}
where $\widecheck{\mathbf{W}}^{\text{add}}_{0}$ is the ensemble transform matrix computed by taking the additional observations as the observations and the analysis as the background.
Then, if the NWP model is available, the forecast ensemble $\mathbf{X}^{f,\text{add}}_{t\mid0}$ can be obtained by integrating the model from the analysis ensemble $\mathbf{X}^{a,\text{add}}_{0}$.
Note that, as shown in Fig.~\ref{fig:concept_efso_add}, the analysis and forecast ensembles with the superscript ``$\text{add}$'' (and the quantities computed from them) are obtained by assimilating the additional observations into the existing analysis, and hence they also include the effect of the existing observations.
In this case, the observation impact of the additional observations $\mathbf{y}^{o,\text{add}}_{0}$ is expressed as
\begin{equation}
    \left(\Delta e^{2}_{t}\right)^{\text{add}}\approx\left(\mathbf{d}^{o-f,\text{add}}_{0}\right)^{\top}\left(\mathbf{R}^{\text{add}}_{0}\right)^{-1}\mathbf{Y}^{a,\text{add}}_{0}\left(\mathbf{Z}^{f,\text{add}}_{t\mid0}\right)^{\top}\mathbf{C}\left(\mathbf{e}^{f,\text{add}}_{t\mid0}+\mathbf{e}^{f}_{t\mid-6}\right), \label{eq:efso_add_add}
\end{equation}
where $\mathbf{d}^{o-f,\text{add}}_{0}\coloneqq\mathbf{y}^{o,\text{add}}_{0}-H^{\text{add}}\left(\overline{\mathbf{x}}^{f}_{0\mid-6}\right)$ and $\mathbf{Y}^{a,\text{add}}_{0}\coloneqq\mathbf{H}^{\text{add}}\mathbf{Z}^{a,\text{add}}_{0}$; $\mathbf{R}^{\text{add}}_{0}$, $H^{\text{add}}$, and $\mathbf{H}^{\text{add}}$ are the observation error covariance matrix, the observation operator, and its Jacobian for the additional observations, respectively.
In addition, $\mathbf{Z}^{f,\text{add}}_{t\mid0}$ and $\mathbf{e}^{f,\text{add}}_{t\mid0}$ are computed from the forecast ensemble $\mathbf{X}^{f,\text{add}}_{t\mid0}$.
Furthermore, if the existing observations $\mathbf{y}^{o}_{0}$ are also available, the total observation impact of the existing observations $\mathbf{y}^{o}_{0}$ and the additional observations $\mathbf{y}^{o,\text{add}}_{0}$ is expressed as
\begin{equation}
    \Delta e^{2}_{t}+\left(\Delta e^{2}_{t}\right)^{\text{add}}\approx\left(\mathbf{d}^{o-f,\text{all}}_{0}\right)^{\top}\left(\mathbf{R}^{\text{all}}_{0}\right)^{-1}\mathbf{Y}^{a,\text{all}}_{0}\left(\mathbf{Z}^{f,\text{add}}_{t\mid0}\right)^{\top}\mathbf{C}\left(\mathbf{e}^{f,\text{add}}_{t\mid0}+\mathbf{e}^{f}_{t\mid-6}\right), \label{eq:efso_add_all}
\end{equation}
where the quantities with the superscript ``$\text{all}$'' are defined as
\begin{equation}
    \mathbf{d}^{o-f,\text{all}}_{0}\coloneqq\left[\begin{array}{c}
        \mathbf{d}^{o-f}_{0}\\
        \mathbf{d}^{o-f,\text{add}}_{0}
        \end{array}\right],\quad
    \mathbf{R}^{\text{all}}_{0}\coloneqq\left[\begin{array}{cc}
        \mathbf{R}_{0} & \mathbf{O}\\
        \mathbf{O} & \mathbf{R}^{\text{add}}_{0}
        \end{array}\right],\quad
    \mathbf{Y}^{a,\text{all}}_{0}\coloneqq\left[\begin{array}{c}
        \mathbf{H}\\
        \mathbf{H}^{\text{add}}
        \end{array}\right]\mathbf{Z}^{a,\text{add}}_{0}, \label{eq:definitions_all}
\end{equation}
and $\mathbf{O}$ denotes the zero matrix.
Note that the total impact of the existing observations $\Delta e^{2}_{t}$ in Eq.~\eqref{eq:efso_add_all} is not strictly identical to that in Eq.~\eqref{eq:efso_all}, because it is evaluated with the additional observations also assimilated. 
The total contribution of the additional observations $\left(\Delta e^{2}_{t}\right)^{\text{add}}$ in Eq.~\eqref{eq:efso_add_all} is identical to that in Eq.~\eqref{eq:efso_add_add} when localization is not applied.
The observation impact with localization applied in the case of additional observations is formulated in the same manner as above.

\subsubsection{PEFSO for additional observations}
Here, when additional observations $\mathbf{y}^{o,\text{add}}_{0}$ become available, we consider estimating the observation impact without reintegrating the forecast model.
By analogy with Eqs.~(\ref{eq:urda_all}, \ref{eq:analysis_add}), the preemptive forecast $\mathbf{X}^{f,\text{add},\text{pre}}_{t\mid0}$ is obtained as
\begin{equation}
    \mathbf{X}^{f,\text{add}}_{t\mid0}\approx\mathbf{X}^{f,\text{add},\text{pre}}_{t\mid0}=\mathbf{X}^{f}_{t\mid0}\widecheck{\mathbf{W}}^{\text{add}}_{0}.
\end{equation}
Accordingly, using the preemptive forecast, Eq.~\eqref{eq:efso_add_add} can be expressed as
\begin{align}
    \left(\Delta e^{2}_{t}\right)^{\text{add}}&\approx\left(\mathbf{d}^{o-f,\text{add}}_{0}\right)^{\top}\left(\mathbf{R}^{\text{add}}_{0}\right)^{-1}\mathbf{Y}^{a,\text{add}}_{0}\left(\mathbf{Z}^{f,\text{add},\text{pre}}_{t\mid0}\right)^{\top}\mathbf{C}\left(\mathbf{e}^{f,\text{add},\text{pre}}_{t\mid0}+\mathbf{e}^{f}_{t\mid-6}\right) \notag\\
    &=\left(\mathbf{d}^{o-f,\text{add}}_{0}\right)^{\top}\left(\mathbf{R}^{\text{add}}_{0}\right)^{-1}\mathbf{Y}^{a,\text{add}}_{0}\left(\mathbf{Z}^{f}_{t\mid0}\mathbf{W}^{\text{add}}_{0}\right)^{\top}\mathbf{C}\left(2\mathbf{e}^{f}_{t\mid-6}+\mathbf{Z}^{f}_{t\mid0}\mathbf{w}^{\text{add}}_{0}\right). \label{eq:pefso_add}
\end{align}
The counterpart of Eq.~\eqref{eq:efso_add_all} and these equations with localization can be formulated in the same manner.

\subsubsection{Ensemble forecast update for additional observations}
We consider obtaining the ensemble forecast in which the additional observations that remain after denying those estimated to be detrimental are assimilated.
Hereafter, among the additional observations $\mathbf{y}^{o,\text{add}}_{0}$, we denote by $\mathbf{y}^{o,\text{add-deny}}$ the subset that is denied as detrimental, and by $\mathbf{y}^{o,\text{add-sel}}$ the remaining subset that is selected as beneficial.
The analysis ensemble $\mathbf{X}^{a,\text{add-sel}}_{0}$ obtained by denying the detrimental observations $\mathbf{y}^{o,\text{add-deny}}$ and assimilating the selected observations $\mathbf{y}^{o,\text{add-sel}}$ is then given by
\begin{equation}
    \mathbf{X}^{a,\text{add-sel}}_{0}\approx\mathbf{X}^{a}_{0}\widecheck{\mathbf{W}}^{\text{add-sel}}_{0}, \label{eq:analysis_add_qc}
\end{equation}
where $\widecheck{\mathbf{W}}^{\text{add-sel}}_{0}$ is the ensemble transform matrix computed by taking the selected additional observations as the observations and the analysis as the background.
The forecast ensemble $\mathbf{X}^{f,\text{add-sel}}_{k\mid0}\; (k > 0)$ is then obtained by reintegrating the forecast model from this analysis ensemble $\mathbf{X}^{a,\text{add-sel}}_{0}$.
However, an environment in which the forecast model can be reintegrated is often unavailable.
Therefore, to update the ensemble forecast more conveniently, we introduce two approaches for obtaining a preemptive forecast $\mathbf{X}^{f,\text{add-sel},\text{pre}}_{k\mid0}$ without reintegrating the forecast model.

The first approach is based on the URDA formulation, as in PEFSO. 
That is, the ensemble transform matrix $\widecheck{\mathbf{W}}^{\text{add-sel}}_{0}$ (i.e., Kalman gain $\mathbf{K}$) is recomputed from the selected additional observations $\mathbf{y}^{o,\text{add-sel}}$, and the preemptive forecast $\mathbf{X}^{f,\text{add-sel},\text{pre}}_{k\mid0}$ is obtained as
\begin{equation}
    \mathbf{X}^{f,\text{add-sel}}_{k\mid0}\approx\mathbf{X}^{f,\text{add-sel},\text{pre}}_{k\mid0}=\mathbf{X}^{f}_{k\mid0}\widecheck{\mathbf{W}}^{\text{add-sel}}_{0}. \label{eq:preemptive_forecast_add_pre_urda}
\end{equation}

The second approach builds upon the approximation proposed by \citet{ota_2013a,hotta_2017,chen_2019}, in which only the innovation is modified. 
The preemptive forecast $\mathbf{X}^{f,\text{add-sel},\text{pre}}_{k\mid0}$ is then given by
\begin{equation}
    \mathbf{X}^{f,\text{add-sel}}_{k\mid0}\approx\mathbf{X}^{f,\text{add-sel},\text{pre}}_{k\mid0}=\mathbf{X}^{f,\text{add},\text{pre}}_{k\mid0}-\mathbf{Z}^{f}_{k\mid0}\mathbf{w}^{\text{add-deny}}_{0}\mathbf{1}^{\top}, \label{eq:preemptive_forecast_add_pre_innovation}
\end{equation}
where $\mathbf{w}^{\text{add-deny}}_{0}\coloneqq\widetilde{\mathbf{P}}^{a,\text{add}}_{0}\left(\mathbf{Y}^{a}_{0}\right)^{\top}\left(\mathbf{R}^{\text{add}}_{0}\right)^{-1}\mathbf{d}^{o-a,\text{add-deny}}_{0}$, and $\mathbf{d}^{o-a,\text{add-deny}}_{0}$ denotes the innovation of the additional observations with the analysis as the background, in which all components other than those corresponding to the rejected subset of observations are set to zero.
The difference between Eq.~\eqref{eq:preemptive_forecast_add_pre_urda} and Eq.~\eqref{eq:preemptive_forecast_add_pre_innovation} is that the former updates the full ensemble, including the perturbations, by recomputing the ensemble transform matrix whereas the latter updates only the ensemble mean by modifying the innovation alone. 
Therefore, Eq.~\eqref{eq:preemptive_forecast_add_pre_urda} is expected to be more accurate. 
On the other hand, Eq.~\eqref{eq:preemptive_forecast_add_pre_innovation} should be less expensive, since only the innovation needs to be modified provided that the relevant quantities have been stored in advance.
    \section{Experimental design}\label{sec:experiment}
\subsection{The Lorenz 96 model}
In this study, we conduct a series of idealized experiments on observation impact estimation and forecast error quantification using the Lorenz 96 model, which has been widely used in theoretical data assimilation studies \citep[e.g.,][]{anderson_2001,whitaker_2002a,ott_2004,kalnay_2012}.
The equations of the Lorenz 96 model are given by
\begin{equation}
    \frac{\mathrm{d}x_{i}}{\mathrm{d}t}=\left(x_{i+1}-x_{i-2}\right)x_{i-1}-x_{i}+F\quad i=1,\dots,n,
\end{equation}
where the model dimension is $n=40$ with cyclic boundary conditions.
The forcing term is set to $F=8.0$, which is a standard value yielding chaotic behavior.
One time step corresponds to a non-dimensional time of $\mathrm{d}t=0.05$, and the model is integrated using the fourth-order Runge--Kutta scheme.
For $F=8.0$, following the discussion of the error doubling time by \citet{lorenz_1998a}, we conventionally treat $\mathrm{d}t=0.05$ as corresponding to $6$ hours.

\subsection{Observation impact estimation for existing observations} \label{sec:experiment_obs_impact_full}
We investigate the influence of the approximations additionally introduced in PEFSO relative to EFSO in terms of the ensemble size, the reference state, and the verification time, using observations at every grid point.
An observing system simulation experiment (OSSE) is conducted to generate the analyses and forecasts used for the observation impact estimation. 
First, after a one-year spin-up, a nature run of 10 years and 30 days is produced. 
Observations at all grid points are then generated by adding Gaussian noise with zero mean and a standard deviation of $0.5$ to the nature run. 
Only at the 10th grid point, however, the observation is generated with a different standard deviation of $2$, so as to represent a detrimental observation. 
The observation error covariance matrix $\mathbf{R}$ used in the assimilation process is assumed to be diagonal, and all of its diagonal elements, including that associated with the observation at the 10th grid point, correspond to a standard deviation of $0.5$. 
The observation at the 10th grid point is therefore assumed to be a detrimental observation that is statistically inconsistent with the data assimilation system. 
This setup is designed to examine whether such a detrimental observation can be correctly detected by EFSO or PEFSO.
As the data assimilation method, the ETKF with an ensemble size of $40$ is employed as an idealized case, and the LETKF with an ensemble size of $10$ as a more realistic one. 
Multiplicative inflation is applied, in which the forecast ensemble perturbations $\delta\mathbf{X}^{f}$ are multiplied by $1.02$ for the ETKF and by $1.04$ for the LETKF. 
For the LETKF, R-localization with a Gaussian function is applied, with a localization scale of $6$.
Assimilation--forecast cycles are performed for 10 years and 30 days with an interval of $\mathrm{d}t=0.05$. 
To avoid the use of poor-quality analyses, the first 30 days are discarded from all data, including the nature run and the observations, and the remaining 10 years are used for the observation impact estimation. 
Furthermore, to ensure independence among the samples, the observation impact is estimated every 24 hours within the 10-year period. 
Consequently, a total of $3651$ observation impact estimations are performed, and the statistical evaluation is based on their time average.
Both the nature run and the analysis are examined as the reference state used in the observation impact estimation. 
The verification time is examined for lead times of 6, 24, 48, 72, 96, 120, 144, and 168 hours (i.e., 0.25, 1, 2, 3, 4, 5, 6, and 7 days). 
The identity matrix is used for $\mathbf{C}$.
When the data assimilation method is the LETKF with an ensemble size of $m=10$, localization is required for EFSO and PEFSO; accordingly, the advective localization of \citet{kalnay_2012} is applied. 
The localization center is shifted by $0.6$ grid points as the verification time extends by one day. 
For PEFSO, the same advective localization is likewise applied to $\mathbf{w}^{\text{loc}}_{0,j}$ and $\mathbf{W}^{\text{loc}}_{0,j}$ in Eq.~\eqref{eq:pefso_all_loc}.

\subsection{Observation impact estimation for additional observations} \label{sec:experiment_obs_impact_add}
We conduct observation impact estimations in a setting where additional observations become available in addition to the existing observations.
Among the 40 grid points, the first 32 points are treated as existing observations, and an OSSE is performed in advance so that analyses and forecasts are available.
The OSSE setting is basically the same as that of section \ref{sec:experiment_obs_impact_full}. 
One difference, however, is that here all existing observations are generated with Gaussian noise of standard deviation $0.5$, and all diagonal elements of the observation error covariance matrix $\mathbf{R}$ likewise correspond to a standard deviation of $0.5$.
That is, no detrimental observations are assumed among the existing observations.
In addition, to reflect a more realistic setting, only the LETKF with an ensemble size of $10$ members is used as the data assimilation method in this experiment. 
The multiplicative inflation parameter is set to $1.02$, and the localization scale for R-localization is set to $6$.

Additional observations are then assumed to become available at the 33rd through 40th points, and the observation impact is estimated.
As with the existing observations, the additional observations are basically generated with Gaussian noise of standard deviation $0.5$, and all diagonal elements of the observation error covariance matrix $\mathbf{R}^{\text{add}}$ likewise correspond to a standard deviation of $0.5$.
Only the observations at grid points 35 and 38, however, are treated as detrimental observations and are generated using Gaussian noise of standard deviation $2$. 
Two detrimental observations are prepared here in order to render the differences in forecast error among the various conditions clearly distinguishable in the experiment of forecast error quantification presented later in section \ref{sec:experiment_quant_fcst_error}.
The assimilation of the additional observations is performed following Eq.~\eqref{eq:analysis_add}. 
For the reference state, only the analysis is adopted as a more realistic choice, since the experiment in section \ref{sec:experiment_obs_impact_full} shows that there is essentially no substantial difference between the truth and the analysis.
In this experiment, we consider both the case in which the existing observations are also available and the case in which only the additional observations are available.
When the existing observations are also available, the observation impacts at all points are estimated, whereas when only the additional observations are available, the impacts of the additional observations alone are estimated.
In doing so, we investigate how the availability of the existing observations affects the results.

\subsection{Ensemble forecast update for additional observations} \label{sec:experiment_quant_fcst_error}
Based on the observation impact estimation in section \ref{sec:experiment_obs_impact_add}, we deny the observations detected as statistically detrimental, obtain ensemble forecasts by assimilating the remaining additional observations, and compare their forecast errors.
In this experiment, the five analysis and forecast ensembles listed in Table \ref{table:fcst_error_quant} are compared in order to examine the effects of the additional observations and of the denied ones.
Table \ref{table:fcst_error_quant}a presents the analysis and forecast update equations for each of them, and Table \ref{table:fcst_error_quant}b summarizes their conceptual differences.
The five approaches are characterized by three aspects: whether the nonlinear forecast model is used to compute the forecast ensemble, which of the additional observations are assimilated, and whether the ensemble transform matrix is recomputed for the assimilated additional observations.
NO-ADD refers to the analysis ensemble $\mathbf{X}^{a}_{0}$ obtained without assimilating the additional observations $\mathbf{y}^{o,\text{add}}_{0}$, together with the forecast ensemble $\mathbf{X}^{f}_{k\mid0}$ integrated from that analysis ensemble with the forecast model, and it serves as the baseline against which the other four approaches are evaluated.
ADD-ALL refers to the analysis ensemble $\mathbf{X}^{a,\text{add}}_{0}$ obtained by assimilating all of the additional observations $\mathbf{y}^{o,\text{add}}_{0}$, together with the forecast ensemble $\mathbf{X}^{f,\text{add}}_{k\mid0}$ integrated from that analysis ensemble with the forecast model.
Since ADD-ALL does not deny the two detrimental observations, it may perform worse than NO-ADD.
ADD-SEL refers to the analysis ensemble $\mathbf{X}^{a,\text{add-sel}}_{0}$ obtained by assimilating only the additional observations $\mathbf{y}^{o,\text{add-sel}}$ selected as beneficial by the observation impact estimation, together with the forecast ensemble $\mathbf{X}^{f,\text{add-sel}}_{k\mid0}$ integrated from that analysis ensemble with the forecast model.
Among these five approaches, ADD-SEL is expected to achieve the best forecast accuracy.
ADD-SEL-PRE serves as an approximation to ADD-SEL using the preemptive forecast and is divided into the following two approaches.
ADD-SEL-PRE-TR (TR: ensemble transform matrix recomputation) recomputes the ensemble transform matrix $\widecheck{\mathbf{W}}^{\text{add-sel}}_{0}$ for the selected additional observations, and therefore employs an analysis ensemble $\mathbf{X}^{a,\text{add-sel}}_{0}$ identical to that of ADD-SEL.
It differs from ADD-SEL, however, in obtaining the preemptive forecast $\mathbf{X}^{f,\text{add-sel},\text{pre}}_{k\mid0}$ from Eq.~\eqref{eq:preemptive_forecast_add_pre_urda} instead of integrating the forecast model.
On the other hand, ADD-SEL-PRE-ID (ID: innovation denial of the detrimental observations) likewise obtains the preemptive forecast without integrating the forecast model, but it differs from ADD-SEL-PRE-TR in that the ensemble transform matrix is not recomputed.
Instead, the analysis ensemble $\mathbf{X}^{a,\text{add-sel}}_{0}$ and the preemptive forecast $\mathbf{X}^{f,\text{add-sel},\text{pre}}_{k\mid0}$ are updated through Eq.~\eqref{eq:preemptive_forecast_add_pre_innovation} by eliminating the contribution of the innovations of the denied observations.
As for the observation impact estimation method, EFSO is used for ADD-SEL under the assumption that the forecast model is available, whereas PEFSO is used for both ADD-SEL-PRE approaches.
Since both of them correctly detect the additional observations at the 35th and 38th grid points as detrimental, no difference arises in this experiment from the choice of the observation impact estimation method.
That is, the differences in forecast error between ADD-SEL and the two ADD-SEL-PRE approaches depend solely on how the analysis and forecast ensembles are updated.

The forecast lead time used for quantifying the forecast error is set to 7 days.
For the verification time in the observation impact estimation, 4 days is selected here as a representative case. 
This is because the additional observations at the 35th and 38th grid points are detected as detrimental regardless of the verification time, and this choice therefore has little influence on the results.
For the localization applied to $\widecheck{\mathbf{W}}^{\text{add-sel}}_{0}$ used in ADD-SEL-PRE-TR and to $\mathbf{w}^{\text{add-deny}}_{0}$ used in ADD-SEL-PRE-ID, advective localization should ideally be employed. 
However, updating it for every forecast lead time $k$ is computationally inefficient, and therefore standard R-localization is used in this experiment.

\begin{table*}[p]
    \centering
    \caption{Summary of the five analysis and forecast ensembles compared in the experiments on ensemble forecast update: (a) analysis and forecast update equations and (b) conceptual differences.}
    \label{table:fcst_error_quant}
    \vspace{0.5em}

    {\renewcommand{\arraystretch}{1.8}%
    \small
    \begin{tabular*}{\textwidth}{@{\extracolsep{\fill}}lll@{}}
        \multicolumn{3}{c}{(a)} \\[0.3em]
        \toprule
        & \multicolumn{1}{c}{Analysis ensemble} & \multicolumn{1}{c}{Forecast ensemble} \\
        \midrule
        NO-ADD (Baseline) &
        $\mathbf{X}^{a}_{0}=\mathbf{X}^{f}_{0\mid-6}\widecheck{\mathbf{W}}_{0}$ &
        $\mathbf{X}^{f}_{k\mid0}=\left[\dots,M_{k\mid0}\left(\mathbf{x}^{a\left(i\right)}_{0}\right),\dots\right]$ \\
        ADD-ALL &
        $\mathbf{X}^{a,\text{add}}_{0}=\mathbf{X}^{a}_{0}\widecheck{\mathbf{W}}^{\text{add}}_{0}$ &
        $\mathbf{X}^{f,\text{add}}_{k\mid0}=\left[\dots,M_{k\mid0}\left(\mathbf{x}^{a,\text{add}\left(i\right)}_{0}\right),\dots\right]$ \\
        ADD-SEL &
        $\mathbf{X}^{a,\text{add-sel}}_{0}=\mathbf{X}^{a}_{0}\widecheck{\mathbf{W}}^{\text{add-sel}}_{0}$ &
        $\mathbf{X}^{f,\text{add-sel}}_{k\mid0}=\left[\dots,M_{k\mid0}\left(\mathbf{x}^{a,\text{add-sel}\left(i\right)}_{0}\right),\dots\right]$ \\
        ADD-SEL-PRE-TR &
        $\mathbf{X}^{a,\text{add-sel}}_{0}=\mathbf{X}^{a}_{0}\widecheck{\mathbf{W}}^{\text{add-sel}}_{0}$ &
        $\mathbf{X}^{f,\text{add-sel},\text{pre}}_{k\mid0}=\mathbf{X}^{f}_{k\mid0}\widecheck{\mathbf{W}}^{\text{add-sel}}_{0}$ \\
        ADD-SEL-PRE-ID &
        $\mathbf{X}^{a,\text{add-sel}}_{0}=\mathbf{X}^{a,\text{add}}_{0}-\mathbf{Z}^{a}_{0}\mathbf{w}^{\text{add-deny}}_{0}\mathbf{1}^{\top}$ &
        $\mathbf{X}^{f,\text{add-sel},\text{pre}}_{k\mid0}=\mathbf{X}^{f,\text{add},\text{pre}}_{k\mid0}-\mathbf{Z}^{f}_{k\mid0}\mathbf{w}^{\text{add-deny}}_{0}\mathbf{1}^{\top}$ \\
        \bottomrule
    \end{tabular*}}

    \vspace{1.5em}

    {\renewcommand{\arraystretch}{1.3}%
    \small
    \begin{tabular*}{\textwidth}{@{\extracolsep{\fill}}lccc@{}}
        \multicolumn{4}{c}{(b)} \\[0.3em]
        \toprule
        & Use of nonlinear & Assimilation of additional & Recomputation of ensemble \\
        & forecast model & observations & transform matrix \\
        \midrule
        NO-ADD (Baseline) & Yes & ---                & ---                      \\
        ADD-ALL           & Yes & All         & Yes                     \\
        ADD-SEL            & Yes & Selected & Yes                     \\
        ADD-SEL-PRE-TR     & No  & Selected & Yes                     \\
        ADD-SEL-PRE-ID     & No  & Selected & No                      \\
        \bottomrule
    \end{tabular*}}
\end{table*}
    \section{Result}\label{sec:result}
\subsection{Observation impact estimation for existing observations} \label{sec:result_obs_impact_full}
\begin{figure}[p]
    \centering
    \includegraphics[width=16cm]{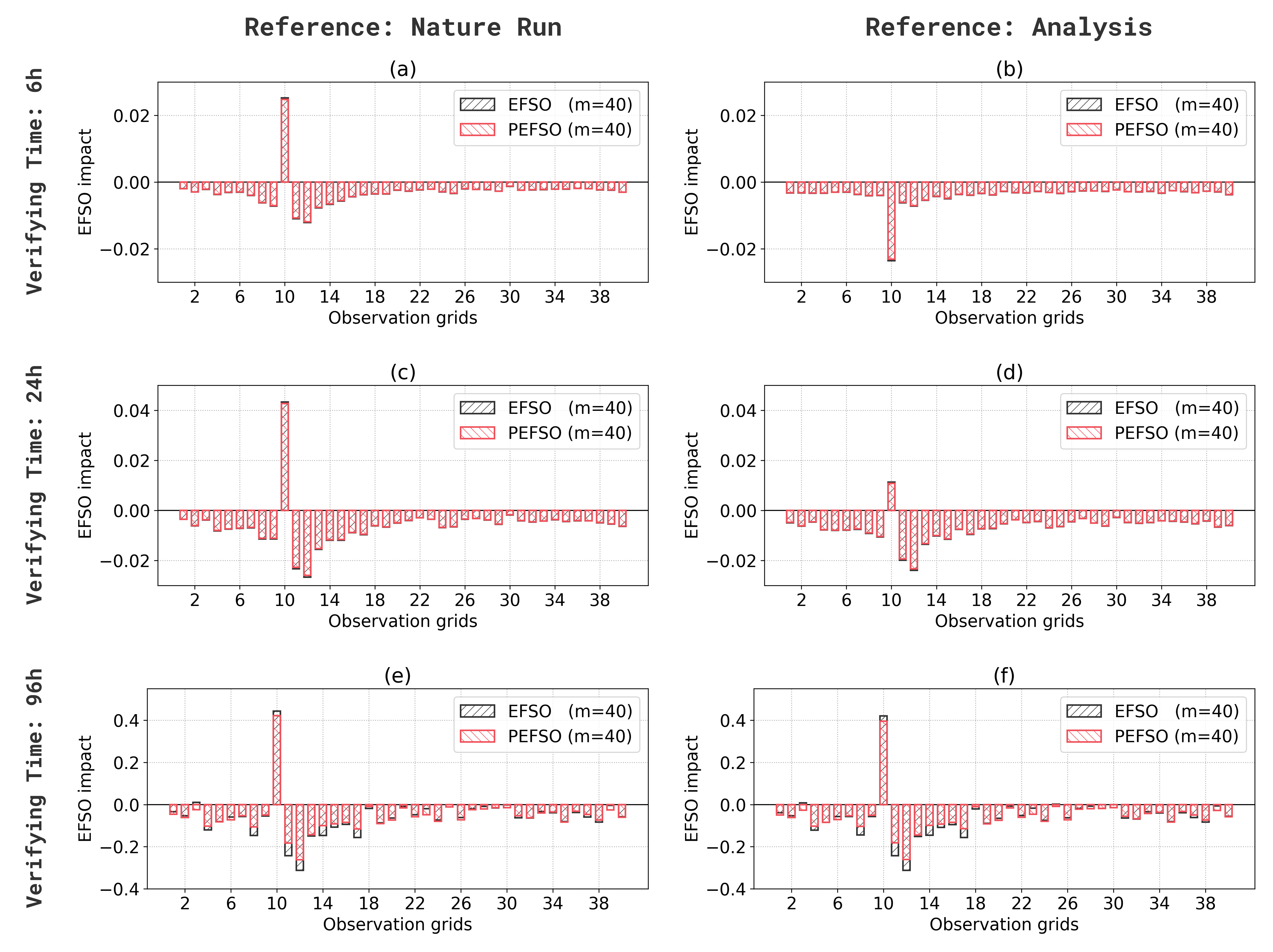}
    \caption{Observation impact of each observation grid point with all observations, for an ensemble size of $m=40$ without localization. 
    The verification times are 6 h in (a) and (b), 24 h in (c) and (d), and 96 h in (e) and (f). 
    The reference state is the nature run in (a), (c), and (e), and the analysis in (b), (d), and (f). 
    Black hatched bars denote the observation impact estimated by EFSO, and red hatched bars that estimated by PEFSO.}
    \label{fig:efso_obs_grid_orig_etkf}
\end{figure}

Figure \ref{fig:efso_obs_grid_orig_etkf} shows the observation impact at each grid point for each verification time and reference state, with an ensemble size of $m=40$ and no localization applied.
Although no reintegration of the forecast model is performed, PEFSO estimates observation impacts comparable to those of EFSO. 
However, at a verification time of 96 h, the discrepancy between the EFSO and PEFSO observation impacts is larger than at 6 h and 24 h.
At the 10th grid point for a verification time of 6 h, the impact is positive when the nature run is used as the reference state, indicating that the detrimental observation is detected as expected. 
In contrast, when the analysis is used as the reference state, the impact at the 10th grid point is negative for both EFSO and PEFSO. 
At a verification time of 24 h, the impact at the 10th grid point is positive even when the analysis is used as the reference state, although its magnitude is smaller than that obtained with the nature run as the reference state.
At a verification time of 96 h, the observation impacts obtained with the analysis as the reference state become comparable to those obtained with the nature run for all observations, including the 10th grid point.

\begin{figure}[p]
    \centering
    \includegraphics[width=16cm]{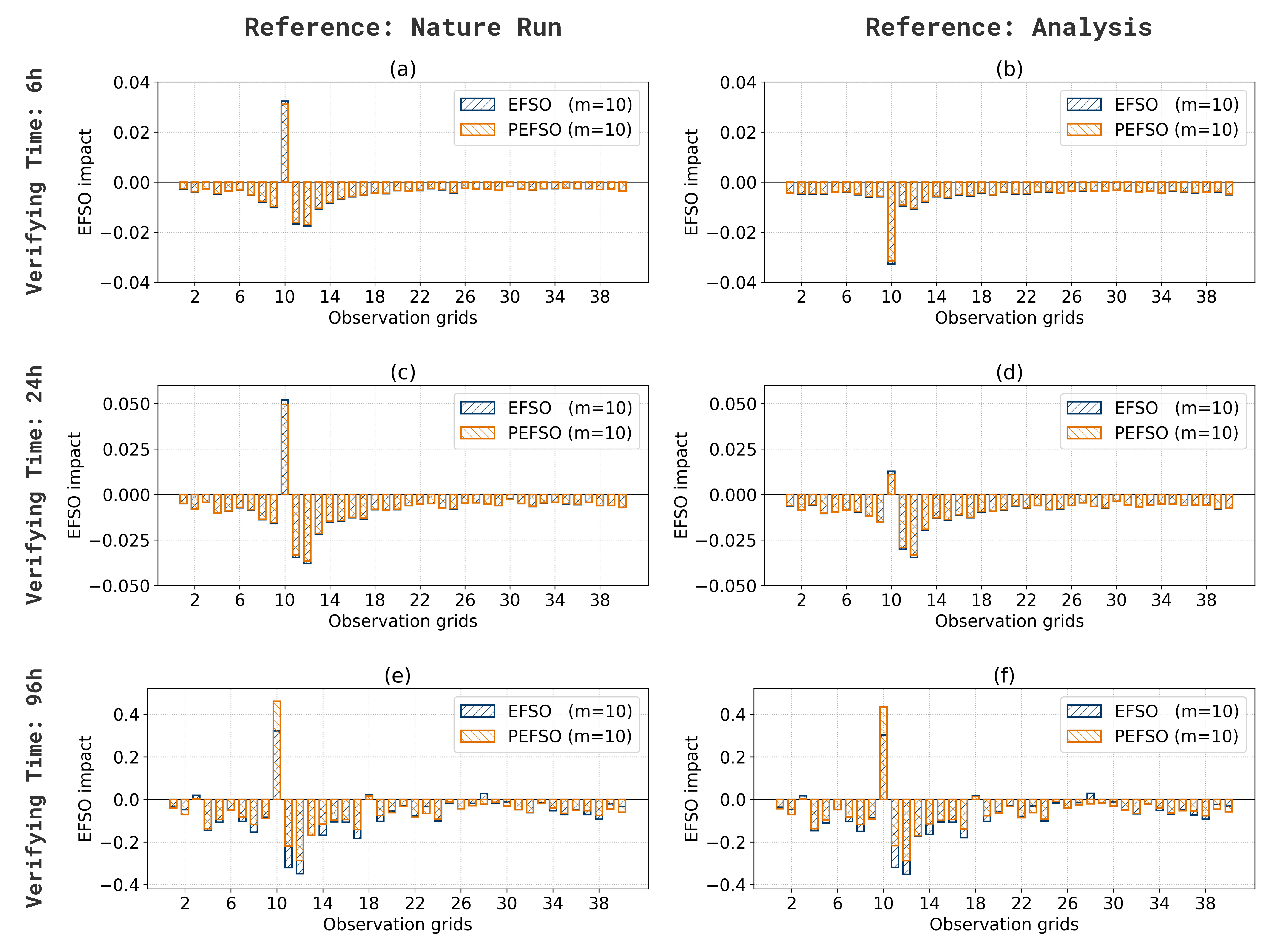}
    \caption{As in Figure \ref{fig:efso_obs_grid_orig_etkf}, but for an ensemble size of $m=10$ with localization applied. 
    Blue hatched bars denote the observation impact estimated by EFSO, and yellow hatched bars that estimated by PEFSO.}
    \label{fig:efso_obs_grid_orig_letkf}
\end{figure}

Figure \ref{fig:efso_obs_grid_orig_letkf} shows the observation impact at each grid point for each verification time and reference state, with an ensemble size of $m=10$ and localization applied.
Even under the more realistic condition of a small ensemble size, PEFSO estimates observation impacts similar to those of EFSO, and the basic tendencies are the same as in Fig.~\ref{fig:efso_obs_grid_orig_etkf}. 
However, the discrepancy between the EFSO and PEFSO observation impacts is slightly larger than in that case, particularly at a verification time of 96 h.

\begin{figure}[p]
    \centering
    \includegraphics[width=14cm]{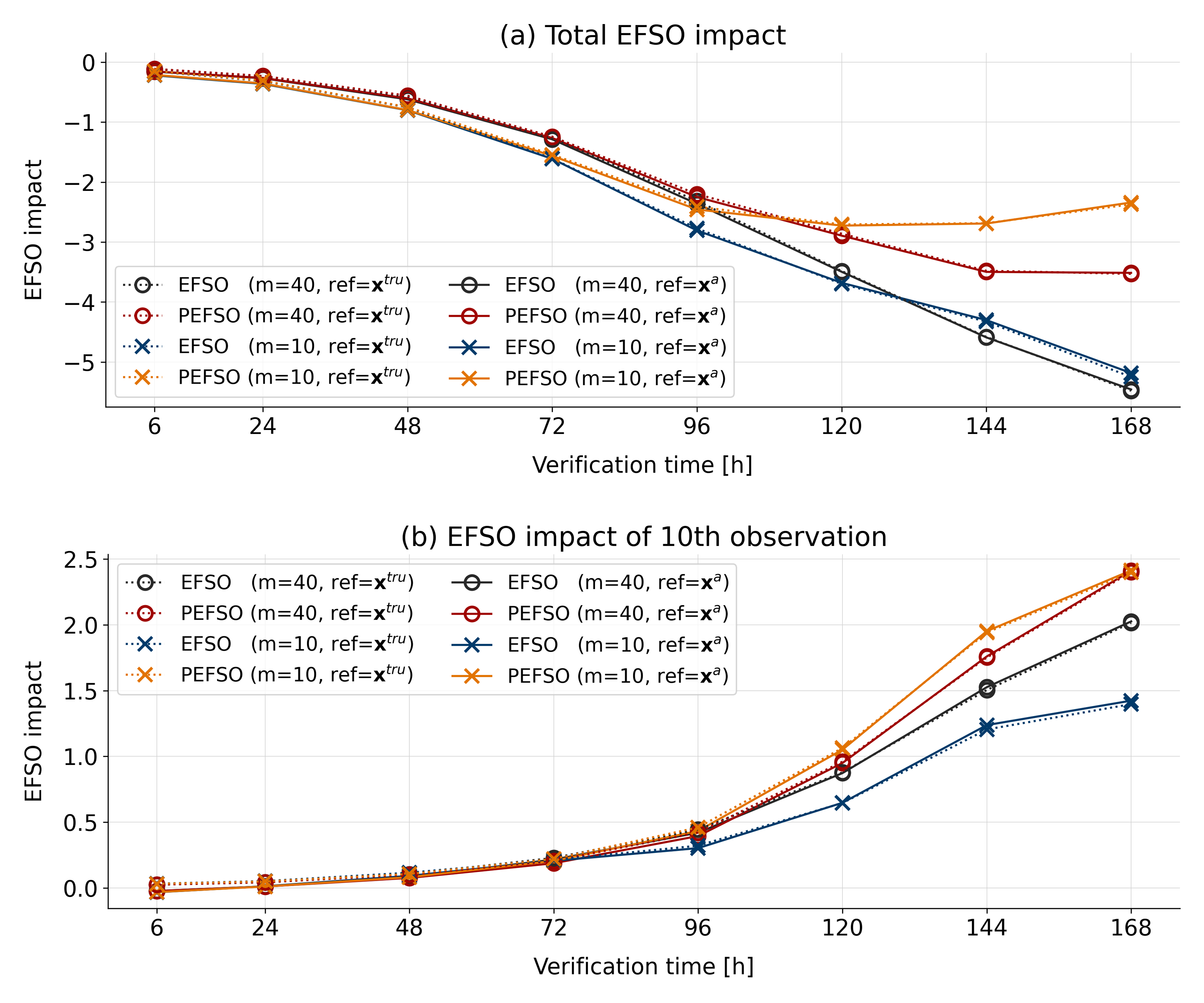}
    \caption{Observation impact against verification time under each condition: (a) total observation impact of all observations, and (b) observation impact of the 10th observation grid point. 
    Black and red lines denote EFSO and PEFSO, respectively, for $m=40$ without localization; blue and yellow lines denote the same for $m=10$ with localization.
    Dashed and solid lines correspond to the nature run and the analysis as the reference state, respectively.}
    \label{fig:efso_verify_time_orig}
\end{figure}

Figure \ref{fig:efso_verify_time_orig} summarizes the observation impacts for each condition at different verification times.
Regarding the estimation method of the observation impact, EFSO and PEFSO show similar observation impacts under all conditions up to a verification time of 96 h, but they gradually diverge at longer verification times. 
For the observation impact at the 10th grid point, PEFSO tends to show larger values than EFSO beyond a verification time of 96 h. 
Accordingly, the same tendency is found for the total observation impact.
Focusing on the ensemble size, the discrepancy between EFSO and PEFSO beyond 96 h tends to be larger for $m=10$ than for $m=40$.

\subsection{Observation impact estimation for additional observations} \label{sec:result_obs_impact_add}
\begin{figure}[p]
    \centering
    \includegraphics[width=13cm]{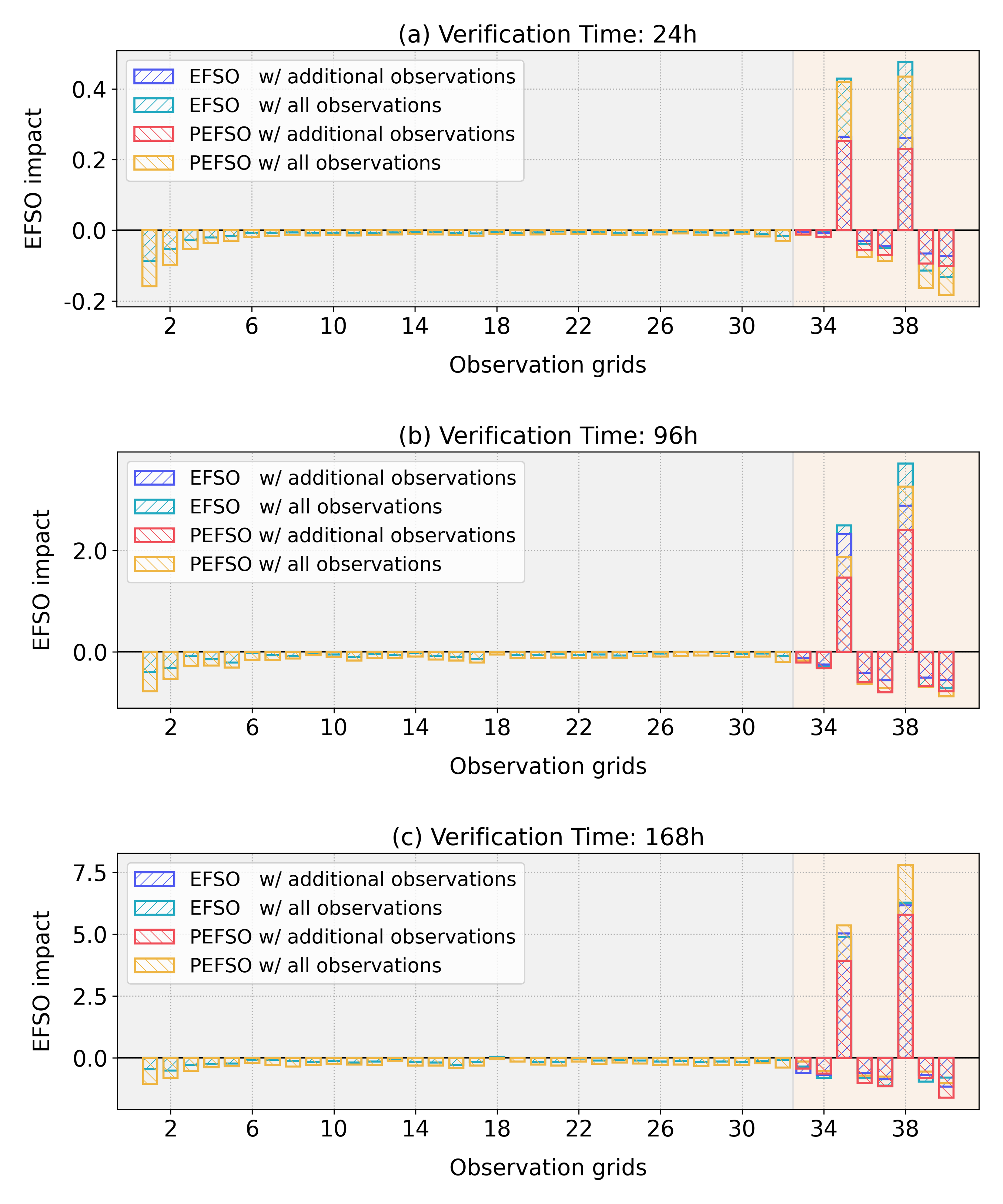}
    \caption{Observation impact of each observation grid point with only the additional observations and with all observations including the existing ones.
    The ensemble size is $m=10$, localization is applied, and the analysis is used as the reference state.
    The verification times are (a) 24 h, (b) 96 h, and (c) 168 h. 
    Blue and light blue hatched bars show EFSO for the additional and all observations, respectively; red and yellow bars show the same for PEFSO.}
    \label{fig:efso_obs_grid_add}
\end{figure}

Figure \ref{fig:efso_obs_grid_add} shows the observation impact at each grid point for EFSO and PEFSO when additional observations are available.
Focusing on the observation impacts of the additional observations (i.e., at the 33rd to 40th grid points), the impacts show similar tendencies regardless of the available observations and the estimation method. 
Specifically, the observation impacts at the 35th and 38th grid points are positive, whereas those at the other grid points are negative, as expected. 
This suggests that PEFSO can estimate the impact of additional observations even when only those observations are available and the forecast model cannot be reintegrated.
However, the absolute value of the observation impact at each grid point tends to be larger for all observations than for only the additional observations, especially at the 35th and 38th grid points.
As for the tendency among the observation grid points, the observation impacts of the additional observations are more dominant than those of the existing observations. 
In addition, the absolute values of the observation impacts tend to be larger on the 40th grid point side (hereafter, downstream) than on the 33rd grid point side (hereafter, upstream).

\begin{figure}[p]
    \centering
    \includegraphics[width=14cm]{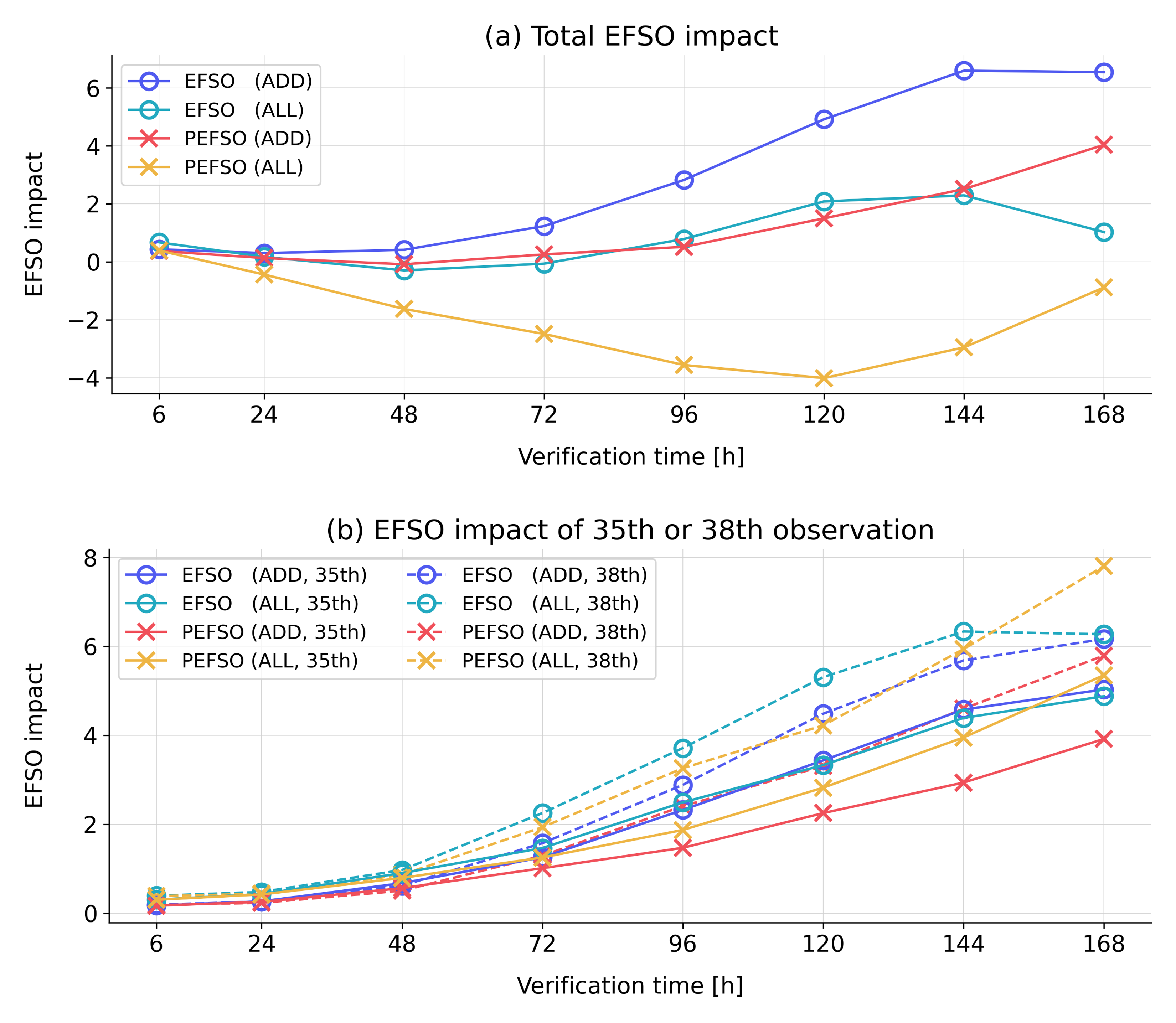}
    \caption{Observation impact estimated by EFSO and PEFSO against verification time with only the additional observations and with all observations including the existing ones. 
    The ensemble size is $m=10$, localization is applied, and the analysis is used as the reference state.
    (a) Total observation impact. Blue and light blue lines show EFSO with only the additional observations and with all observations, respectively; red and yellow lines show the same for PEFSO. 
    (b) Observation impact of the 35th or 38th observation grid point. The colors are as in (a). 
    Solid lines show the observation impact of the 35th grid point, and dashed lines that of the 38th grid point.}
    \label{fig:efso_verify_time_add}
\end{figure}

Figure \ref{fig:efso_verify_time_add} summarizes the observation impacts for each condition at different verification times.
Focusing on the total observation impact in Fig.~\ref{fig:efso_verify_time_add}a, for the same estimation method, the total impact tends to be smaller for all observations than for only the additional observations, because the existing observations are basically beneficial.
However, with only the additional observations, the order of magnitude of the total impact is not substantially different from that for all 40 grid points, although it is the total over only eight grid points.
This suggests that the observation impacts of the additional observations are dominant, as seen in Fig.~\ref{fig:efso_obs_grid_add}.
Next, focusing on the observation impacts at the 35th and 38th grid points in Fig.~\ref{fig:efso_verify_time_add}b, the impacts are positive at all verification times, indicating that the detrimental observations are successfully detected. 
While the impacts are comparable under all conditions up to a verification time of 48 h, they gradually diverge thereafter.
As shown in Fig.~\ref{fig:efso_obs_grid_add}, for the same estimation method, the impacts tend to be larger for all observations than for only the additional observations. 
Furthermore, as also shown in Fig.~\ref{fig:efso_obs_grid_add}, the observation impact at the 38th grid point tends to be larger than that at the 35th grid point throughout.

\subsection{Ensemble forecast update for additional observations} \label{sec:result_quant_fcst_error}
\begin{figure}[p]
    \centering
    \includegraphics[width=16cm]{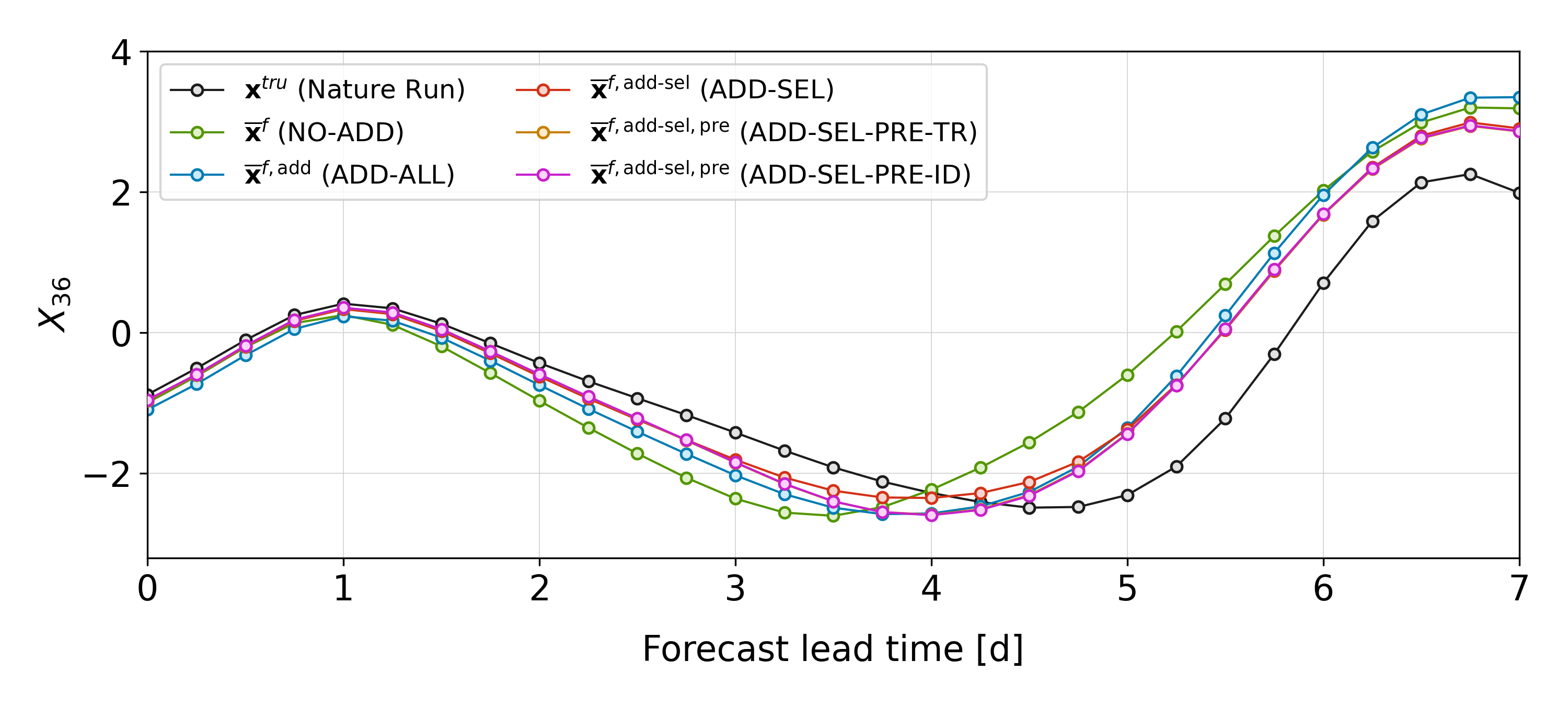}
    \caption{An example of the nature run and the forecast ensemble means of $X_{36}$ at an arbitrarily selected initial time. 
    The black line shows the nature run, the green line NO-ADD, the blue line ADD-ALL, the red line ADD-SEL, the orange line ADD-SEL-PRE-TR, and the pink line ADD-SEL-PRE-ID.}
    \label{fig:qc_forecast}
\end{figure}

Figure \ref{fig:qc_forecast} shows an example of the nature run and the forecasts of $X_{36}$ at an arbitrarily selected initial time.
All forecasts are close to the nature run up to day 1, but they gradually deviate from it thereafter.
The ADD-SEL forecast and both ADD-SEL-PRE forecasts remain relatively closer to the nature run throughout, whereas the NO-ADD and ADD-ALL forecasts are less accurate than these.

\begin{figure}[p]
    \centering
    \includegraphics[width=16cm]{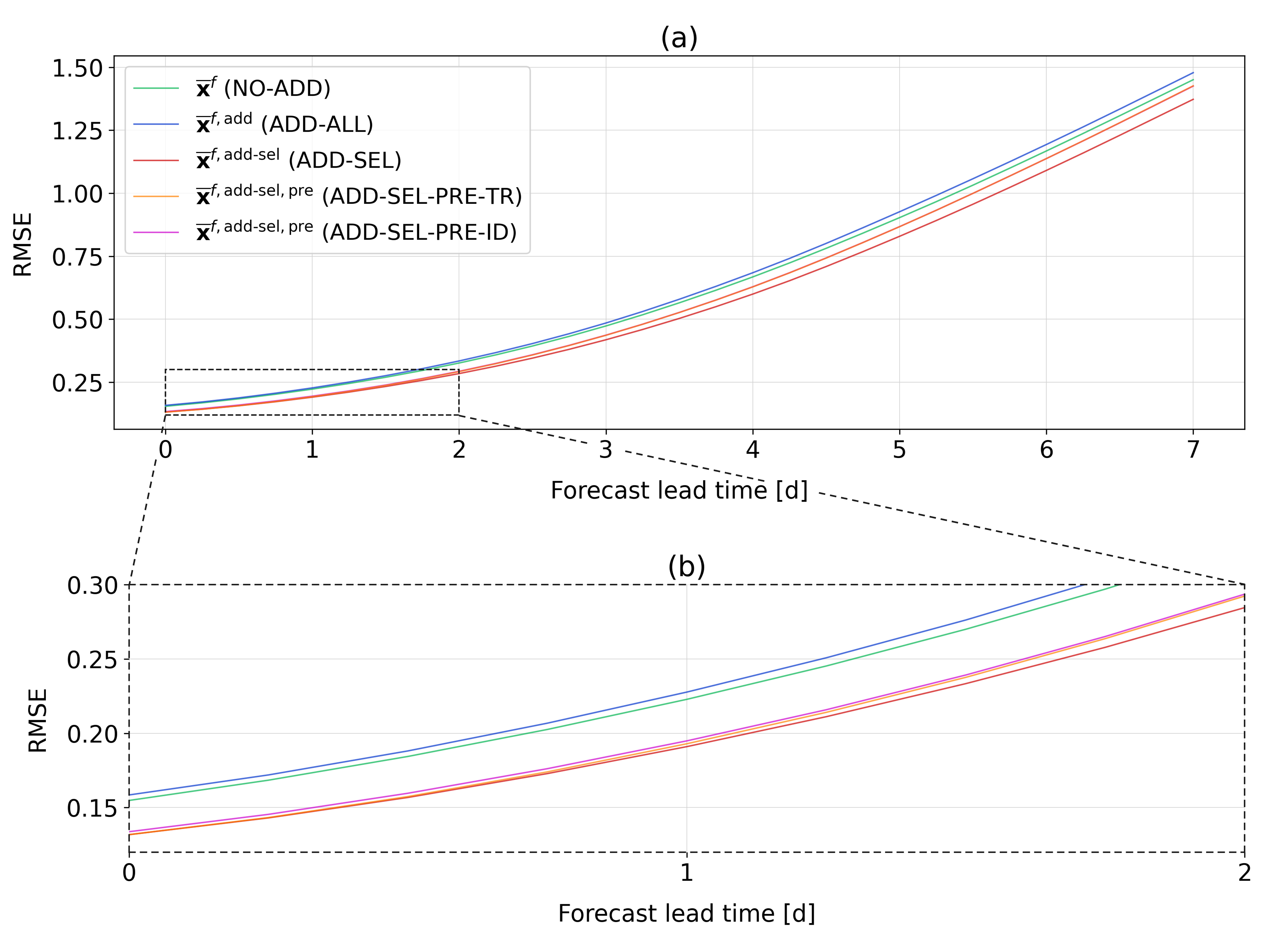}
    \caption{Comparison of the forecast errors. 
    (a) Overview of the 7-day forecast period. 
    (b) Enlarged view of the first 2 days in (a). 
    The green line corresponds to NO-ADD (Baseline) , the blue line ADD-ALL, the red line ADD-SEL, the orange line ADD-SEL-PRE-TR, and the pink line ADD-SEL-PRE-ID.}
    \label{fig:qc_rmse}
\end{figure}

Figure \ref{fig:qc_rmse} shows the statistical forecast errors of the individual forecasts.
Compared with NO-ADD, the forecast error of ADD-ALL is instead degraded because the detrimental observations at the 35th and 38th grid points are also assimilated. 
In contrast, the ADD-SEL forecast and both ADD-SEL-PRE forecasts show substantial reductions in forecast error compared with NO-ADD owing to the removal of the detrimental observations.
Specifically, although the forecast errors of both ADD-SEL-PRE forecasts degrade relative to ADD-SEL as the forecast lead time is extended, they still show an improvement over NO-ADD. 
Moreover, despite no reintegration of the forecast model, both ADD-SEL-PRE forecasts achieve accuracy comparable to that of ADD-SEL, particularly up to a forecast lead time of about 2 days.
Furthermore, a closer look at the forecast errors up to day 2 in Fig.~\ref{fig:qc_rmse}b shows that the analyses of ADD-SEL and ADD-SEL-PRE-TR are identical, whereas the analysis of ADD-SEL-PRE-ID is slightly less accurate. 
Then, as the forecast proceeds, the forecast error of ADD-SEL-PRE-TR gradually approaches that of ADD-SEL-PRE-ID.

    \section{Discussion}\label{sec:discussion}
\subsection{Tendency of the observation impacts}
\subsubsection{Reference state and verification time}
In Figs.~\ref{fig:efso_obs_grid_orig_etkf} and \ref{fig:efso_obs_grid_orig_letkf}, at a verification time of 6 h, the observation impact at the 10th grid point was negative when the self-analysis was used as the reference state, even though it was expected to be positive.
This result can be interpreted in light of the discussion by \citet{hotta_2017}.

When the self-analysis is used as the reference state and the verification time is set to the analysis time, Eq.~\eqref{eq:efso_all} for the total observation impact is expressed as
\begin{equation}
    \Delta e^{2}_{0}=-\left(\mathbf{e}^{f}_{0\mid-6}\right)^{\top}\mathbf{C}\mathbf{e}^{f}_{0\mid-6}.\label{eq:efso_self_ana_all}
\end{equation}
Furthermore, Eq.~\eqref{eq:efso_individual} for the observation impact corresponding to the $l$-th observation $y^{o}_{t,l}$ can be written as
\begin{equation}
    \Delta e^{2}_{0,l}=-\left[\mathbf{K}^{\top}_{0}\mathbf{C}\mathbf{K}_{0}\right]_{ll}\left(\left[\mathbf{d}^{o-f}_{0}\right]_{l}\right)^{2}-\left[\mathbf{d}^{o-f}_{0}\right]_{l}\sum^{p}_{i=1,i\neq l}\left[\mathbf{K}^{\top}_{0}\mathbf{C}\mathbf{K}_{0}\right]_{li}\left[\mathbf{d}^{o-f}_{0}\right]_{i}. \label{eq:efso_self_ana_individual}
\end{equation}
Since $\mathbf{C}$ is positive definite, $\mathbf{K}^{\top}_{0}\mathbf{C}\mathbf{K}_{0}$ is positive semidefinite, and thus the first term is non-positive.
The second term can take either positive or negative values.
Therefore, when the self-analysis is used as the reference state and the verification time is set to the analysis time, the individual observation impacts are not necessarily negative, but Eq.~\eqref{eq:efso_self_ana_all} indicates that negative values tend to appear because the total observation impact is negative.
In addition, Eq.~\eqref{eq:efso_self_ana_individual} indicates that the observation impact tends to be negative particularly when the magnitude of the innovation corresponding to the $l$-th observation $y^{o}_{t,l}$ is larger than those of the other observations.

Indeed, in Fig.~3 of \citet{hotta_2017}, the observation impacts of MODIS winds showed similar spatial patterns for verification times of 6 h and 24 h, whereas a different result was exhibited when the verification time was set to the analysis time, with negative impacts at many locations.
In this study, we also confirmed that the observation impacts are negative at all grid points, and particularly at the 10th grid point, when the verification time is set to the analysis time (not shown).
Furthermore, unlike \citet{hotta_2017}, Figs.~\ref{fig:efso_obs_grid_orig_etkf} and \ref{fig:efso_obs_grid_orig_letkf} show a similar result even at a verification time of 6 h.
This can be explained by the difference in the systems considered between this study and \citet{hotta_2017}.
That is, we infer that a more accurate analysis is obtained in the full-observations experiment with the Lorenz 96 model (section \ref{sec:experiment_obs_impact_full}) than in the realistic NWP system of \citet{hotta_2017}, and that the error growth is therefore slower.
In fact, the observation impact at the 10th grid point in Figs.~\ref{fig:efso_obs_grid_orig_etkf} and \ref{fig:efso_obs_grid_orig_letkf} still differs between the nature run and the analysis as the reference state at a verification time of 24 h and becomes comparable at 96 h, which supports this inference.
Furthermore, Fig.~\ref{fig:efso_verify_time_add}b shows positive observation impacts at the 35th and 38th grid points, which are expected to be detrimental, even at a verification time of 6 h, despite the analysis being used as the reference state.
This can be interpreted as indicating that the error grew sufficiently because the accuracy of the analysis was degraded in the additional-observation experiment (section \ref{sec:experiment_obs_impact_add}), where only 32 existing observations were available.

Therefore, when the self-analysis is used as the reference state and the verification time is close to the analysis time, the individual observation impacts tend to be negative even for observations that can degrade the forecast.
We further argue that the verification time should be determined appropriately, taking into account the error growth rate of the system.

\subsubsection{Observation grid points}
In Figs.~\ref{fig:efso_obs_grid_orig_etkf} and \ref{fig:efso_obs_grid_orig_letkf}, the observation impacts in the vicinity of the detrimental observation at the 10th grid point (i.e., at the 9th and 11th grid points) tend to be larger in absolute value than those at the other grid points, which is consistent with the results of \citet{liu_2008c} (e.g., their Fig.~5).
As discussed by \citet{liu_2008c}, this is presumably because the forecast error variance is relatively large near the grid point with the detrimental observation, so that a greater weight is given to the observations in the assimilation process.

As shown in Fig.~\ref{fig:efso_obs_grid_add}, the observation impacts of the additional observations are more dominant than those of the existing observations.
This tendency can also be explained in the same way: the forecast error variance is relatively large at the grid points where no existing observations were available.

Furthermore, in all the results shown in Figs.~\ref{fig:efso_obs_grid_orig_etkf}, \ref{fig:efso_obs_grid_orig_letkf}, and \ref{fig:efso_obs_grid_add}, the observation impacts tend to be larger in absolute value at the grid points located in the downstream direction (i.e., the direction from $X_{1}$ to $X_{40}$).
For example, comparing the observation impacts at the 35th and 38th grid points in Fig.~\ref{fig:efso_obs_grid_add}, that at the more downstream 38th grid point is markedly larger.
This may be related to the dominance of the group velocity in the downstream direction in the Lorenz 96 model, as described by \citet{lorenz_1998a,kalnay_2012}.
That is, we infer that such a tendency arises because the analysis increment (i.e., the influence of the observations) propagates downstream as the forecast proceeds.
For instance, focusing on the cases in Figs.~\ref{fig:efso_obs_grid_orig_etkf} and \ref{fig:efso_obs_grid_orig_letkf}, the detrimental observation at the 10th grid point affects the forecast downstream of it.
Accordingly, the forecast error variance becomes relatively large on the 11th grid point side, and consequently the absolute value of the observation impact may also become larger.
Similarly, in the case of Fig.~\ref{fig:efso_obs_grid_add}, the forecast at the 33rd grid point benefits from the existing observations immediately upstream near the 32nd grid point, whereas the forecast at the 40th grid point has no such observations close enough upstream.
Hence, the forecast error variance is again relatively larger toward the 40th grid point than toward the 33rd grid point, and consequently the absolute value of the observation impact may also become larger.
This tendency is expected to be particularly pronounced when localization is applied.

\subsection{The effectiveness and limits of PEFSO and ADD-SEL-PRE} 
This study has demonstrated that PEFSO can yield observation impacts comparable to those of EFSO, although no reintegration of the forecast model is performed.
It is particularly promising that, in the additional-observation experiment, PEFSO estimated observation impacts consistent with those of EFSO and successfully detected the detrimental observations even under the more realistic conditions of an ensemble size of $m=10$, the analysis as the reference state, and the availability of the additional observations only.
This is presumably because, although PEFSO imposes a stronger tangent linear approximation, EFSO is itself formulated on the basis of the tangent linear approximation, so that PEFSO also works well as long as this approximation holds sufficiently well.
Conversely, however, as already seen in Figs.~\ref{fig:efso_verify_time_orig} and \ref{fig:efso_verify_time_add}, it should be noted that the PEFSO observation impacts diverge from those of EFSO once the tangent linear approximation no longer holds as the verification time is extended.

Furthermore, in quantifying the forecast error, both ADD-SEL-PRE forecasts also achieved reductions in forecast error comparable to those of the ADD-SEL forecast, again without reintegrating the forecast model.
Specifically, the forecast errors of both ADD-SEL-PRE forecasts are very close to those of the ADD-SEL forecast, particularly up to a forecast lead time of about 2 days, which is a promising result.
However, since both ADD-SEL-PRE forecasts are also based on the tangent linear approximation, it should be noted that their forecast errors diverge from those of the ADD-SEL forecast as the forecast lead time is extended.

A limitation of this study is that, since it is based on an idealized experiment, it remains unclear how effective PEFSO and ADD-SEL-PRE are for the real atmosphere.
In particular, the real atmosphere may exhibit stronger nonlinearity than the Lorenz 96 model, which would make the tangent linear approximation more difficult to hold.
For example, in moist convective processes, the nonlinearity is so strong that the tangent linear approximation is inherently difficult to justify.
Whereas verification times of 6--24 h have commonly been used in previous EFSO studies \citep[e.g.,][]{hotta_2017}, investigating the valid range of verification times for PEFSO and ADD-SEL-PRE is essential for their practical application.
In addition, in this study, the detrimental observations were assigned an error standard deviation four times that used in the observation error covariance matrix in the data assimilation, which is a clearly inconsistent setting.
In reality, however, detrimental observations are not limited to such clearly inconsistent ones.
It thus remains unclear whether PEFSO can correctly detect detrimental observations whose inconsistency is less pronounced as accurately as EFSO.
Moreover, detrimental observations arise not only from an inconsistent error variance but also from various other factors such as biases.
With respect to the forecast model, uncertainties such as model error, which were not investigated in this study, may also affect the estimation of the observation impact and the quantification of the forecast error.
For ADD-SEL-PRE in particular, the difference between ADD-SEL-PRE-TR and ADD-SEL-PRE-ID may become larger in more complex assimilation and forecasting systems, and thus an appropriate choice between them should be made in consideration of the target forecast lead time and the computational cost.
That said, given that EFSO has been shown to work in previous studies in the real atmosphere \citep[e.g.,][]{ota_2013a,sommer_2014}, PEFSO is also expected to estimate comparable observation impacts as long as the tangent linear approximation holds sufficiently well.
To confirm this expectation and to properly address the limitations discussed here, further verification using more realistic NWP models is required toward practical application of PEFSO and ADD-SEL-PRE.
    \section{Conclusion}\label{sec:conclusion}
In this study, assuming a situation in which new observations become available for an existing NWP system, we proposed a method called PEFSO that estimates the observation impact more readily without reintegrating the forecast model.
By imposing a stronger tangent linear approximation than EFSO, PEFSO approximates the observation impact using the existing ensemble forecasts.
Furthermore, we proposed two methods for updating, without reintegration, the ensemble forecast obtained by assimilating the additional observations after denying the detrimental ones.
The first, ADD-SEL-PRE-TR, recomputes the ensemble transform matrix and thereby updates the full ensemble including the perturbations.
The second, ADD-SEL-PRE-ID, modifies only the innovation and updates only the ensemble mean at a lower computational cost.

Through a series of experiments with the Lorenz 96 model, we then examined whether the observation impact can be estimated and the ensemble forecast updated with sufficient accuracy without reintegrating the forecast model.
The results showed that PEFSO can estimate observation impacts comparable to those of EFSO as long as the tangent linear approximation holds sufficiently well.
It is particularly promising that, in the experiment with additional observations, PEFSO estimated observation impacts consistent with those of EFSO and successfully detected the detrimental observations even under the more realistic conditions of an ensemble size of $m=10$, the analysis as the reference state, and the availability of the additional observations only.
In addition, when the additional observations were assimilated after denying the detrimental ones, both ADD-SEL-PRE methods updated the ensemble forecast so as to give forecast errors comparable to those of ADD-SEL, which relies on the forecast model.
In particular, the forecast errors were very close up to a forecast lead time of 2 days, within which the tangent linear approximation is expected to hold well.
On the other hand, when the tangent linear approximation no longer holds as the verification time and the forecast lead time are extended, the results without reintegration depart from those with reintegration.
We therefore conclude that PEFSO and ADD-SEL-PRE work well within the range where the tangent linear approximation is sufficiently valid.

Such estimation of the observation impact and updating of the ensemble forecast without reintegrating the forecast model can serve as a useful tool in the development of new observing methods.
We emphasize, however, that the proposed methods are intended only for preliminary estimation prior to formal evaluation and are not intended to replace conventional approaches for evaluating observation impacts, such as OSE and EFSO.
In addition, since this study is based on idealized experiments, verification using more realistic NWP systems is required for practical application.
    \section*{Acknowledgements}
The authors thank Dr. Atsushi Okazaki and Dr. Kenta Kurosawa of Chiba University for valuable discussions and insightful comments.
This study was supported by the Japan Society for the Promotion of Science (JSPS) through KAKENHI (Grants JP25KJ0729 and JP25H00752), the Japan Science and Technology Agency (JST) Moonshot Research and Development Program (Grant JPMJMS2389), and the Institute for Advanced Academic Research (IAAR) Research Support Program of Chiba University.
    \begingroup
	\normalem
	\bibliographystyle{apalike}
	\bibliography{references.bib}
	\endgroup
\end{document}